\documentclass[12pt]{article}

\usepackage{indentfirst}
\date{%
    \small $^1$School of Computing Science, Simon Fraser University\\
    [2ex]
}

\usepackage{amsmath}
\usepackage{amsthm}
\newtheorem{theorem}{Theorem}
\newtheorem{axiom}{Axiom}
\newtheorem{corollary}{Corollary}
\usepackage{enumitem}
\usepackage{amsfonts}
\usepackage{booktabs}
\usepackage[T1]{fontenc}
\usepackage{array}
\usepackage{xcolor}
\usepackage{xspace}
\definecolor{mydarkblue}{rgb}{0,0.08,0.45}
\usepackage[colorlinks,citecolor=mydarkblue]{hyperref}
\usepackage{todonotes}
\usepackage{natbib}
\usepackage{cleveref}
\Crefname{section}{Sec.}{Secs.}
\crefname{section}{Section}{Sections}
\crefname{table}{Table}{Tables}
\Crefname{table}{Tab.}{Tabs.}
\crefname{figure}{Fig.}{Figs.}
\crefname{theorem}{Theorem}{Theorem}
\crefname{corollary}{Corollary}{Corollary}
\crefname{axiom}{Axiom}{Axiom}
\crefname{appendix}{Appendix}{Appendix}
\theoremstyle{definition}
\newtheorem{definition}{Definition}
\usepackage{etoolbox,chngcntr}

\usepackage{graphicx}
\usepackage[autostyle]{csquotes}

\DeclareRobustCommand\onedot{\futurelet\@let@token\@onedot}
\def\onedot{\ifx\@let@token.\else\null\fi\xspace}

\def\ie{i.e.\onedot}

\providecommand{\keywords}[1]{\textbf{Keywords:} #1}

\usepackage{mathtools}
\usepackage{multirow}
\usepackage{placeins}
\usepackage{array,ragged2e,longtable}
\newcolumntype{P}{>{\RaggedRight\hspace{0pt}}p{\dimexpr(\textwidth-1cm -10\tabcolsep -5\arrayrulewidth)/4\relax}}

\begin{document}
\title{
Ethical Medical Image Synthesis
}
\author{Weina Jin$^1$, Ashish Sinha$^1$, Kumar Abhishek$^1$, Ghassan Hamarneh$^1$
}
\maketitle              %

\begin{abstract}
    The task of ethical \textbf{M}edical \textbf{I}mage \textbf{Syn}thesis (MISyn) is to ensure that the MISyn techniques are researched and developed ethically throughout their entire lifecycle, which is essential to prevent the negative impacts of MISyn. To address the ever-increasing needs and requirements for ethical practice of MISyn research and development, we first conduct a theoretical analysis that identifies the key properties of ethical MISyn and intrinsic limits of MISyn. We identify that synthetic images lack inherent grounding in real medical phenomena, cannot fully represent the training medical images, and inevitably introduce new distribution shifts and biases. 
    Ethical risks can arise from not acknowledging the intrinsic limits and weaknesses of synthetic images compared to medical images, with the extreme form manifested as misinformation of MISyn that substitutes synthetic images for medical images without acknowledgment. The resulting ethical harms include eroding trust in the medical imaging dataset environment and causing algorithmic discrimination towards stakeholders and the public.
    To facilitate collective efforts towards ethical MISyn within and outside the medical image analysis community, we then propose practical supports for ethical practice in MISyn based on the theoretical analysis, including ethical practice recommendations that adapt the existing technical standards, problem formulation, design, and evaluation practice of MISyn to the ethical challenges; and oversight recommendations to facilitate checks and balances from stakeholders and the public. We also present two case studies that demonstrate how to apply the ethical practice recommendations in practice, and identify gaps between existing practice and the ethical practice recommendations.

\end{abstract}
\keywords{Medical image synthesis; AI ethics; Generative AI; Misinformation; Algorithmic discrimination}

\section{Introduction}
Medical image synthesis (MISyn) is the use of computational techniques to generate ``visually realistic and quantitatively accurate images'' in biomedicine~\citep{8305584} based on appropriate inferences
from a given input, such as a source image~\citep{reynaud2023feature},
image modality~\citep{8859286,dalmaz2022resvit}, segmentation mask~\citep{shin2018medical,jin2018ct,abhishek2019mask2lesion,thambawita2022singan,van2024echocardiography,nguyen2024training}, 
image class~\citep{liu2023one}, or a text prompt~\citep{akrout2023diffusion,10566053,bluethgen2024vision,lin2025skingen}.
MISyn appears in a family of medical image analysis (MIA) tasks,
such as medical image-to-image translation~\citep{doi:10.1148/radiol.232471}, image super-resolution~\citep{PRINCE20201},
simulation~\citep{8305584}, data augmentation~\citep{DAYARATHNA2024103046}, 
image registration~\citep{PRINCE20201}, 
and image denoising~\citep{kazerouni2023diffusion}.
The main motivations and objectives of MISyn include generating high-quality datasets~\citep{8305584}, imputing missing values to fulfill the requirements of downstream model learning~\citep{10.1145/3614425}, 
image standardization~\citep{ALAJAJI2024100369}, data augmentation (ibid.), domain adaptation (ibid.), 
 marker removal (ibid.), privacy protection~\citep{ZHANG2024102996}, etc. 
There have been impressive technical advances in MISyn, and many of them utilize state-of-the-art data-driven technologies that are now claimed under the name of artificial intelligence/machine learning (AI/ML)~\citep{Suchman2023,Goodlad2023}. These data-driven generative techniques learn latent data representations from the training data and prior domain knowledge, and generate images by drawing inferences from the probabilistic latent space that maximizes the probabilistic distribution of the images. Representative techniques of generative models include variational autoencoder (VAE)~\citep{kingma2013auto}, generative adversarial network (GAN)~\citep{goodfellow2014generativeadversarialnetworks,8859286}, Gaussian splatting~\citep{10.1007/978-3-031-72089-5_24}, and in recent years, the large pre-trained models~\citep{ZHANG2024102996} and its underpinning models such as diffusion~\citep{KAZEROUNI2023102846} and transformers~\citep{AZAD2024103000}.

Paralleling the technical momentum is the rise of ethically questionable conduct of technical development and their negative social consequences, 
broadly in AI and particularly in generative AI (GenAI), such as algorithmic discrimination~\citep{chunDiscriminatingDataCorrelation2021,browneFeministAICritical2024}, data labor exploitation~\citep{neda_surrogate_2019,gray_ghost_2019,williams_exploited_2022}, widening inequalities~\citep{wigginsHowDataHappened2023}, concentration of power~\citep{crawfordAtlasAIPower2021}, copyright infringement~\citep{GenerativeAIHas}, non-consensual use of data~\citep{AIOpenLetter}, and environmental harms~\citep{JMLR:v21:20-312,li2023makingaithirstyuncovering,crawfordAtlasAIPower2021}. Therefore, understanding and adhering to ethics has become an urgent need and requirement for technical development to ensure that technologies can serve for social good and minimize harms. Conducting ethical and responsible research provides pathways towards rigorous science\footnote{The relationship between ethical and scientific practice can be summarized as: Scientifically rigorous practice is a necessary condition for ethical practice.}~\citep{Andrews2024}. Being ethical is also a prerequisite for conducting responsible research regulated by research ethics~\citep{10.1093/schbul/sbj005}, and a prerequisite for developing medical technologies in clinical settings to comply with existing bioethics~\citep{fda2023} and regulations~\citep{Kong79} in medical practice.

In recent years, there has been a growing body of works that critically analyze the ethics, philosophy, and politics of synthetic data in the general language and vision tasks~\citep{https://doi.org/10.1038/s44319-024-00101-0, Offenhuber2024,Bendel2023,Jacobsen2023,Steinhoff2022,Symons2019,hao2024syntheticdataaichallenges,EthicalConsiderationsRelating} and in medicine~\citep{HASAN2024,ning2024generativeartificialintelligencehealthcare, Chauhan2023}. For MISyn tasks, its ethical investigations and discussions are still in the inception stage, with few works discussing ethical issues of MISyn including misinformation, patient privacy, informed consent, autonomy, algorithmic and data bias, fairness, transparency, accountability, and malicious use~\citep{10.1007/978-3-031-72787-0_17,das2024ethicalframeworkresponsiblefoundational,Paladugu2023,ALAJAJI2024100369,Arora2022,KoohiMoghadam2023}.
Despite these efforts and individual researchers’ and practitioners’ awareness of conducting research and development ethically, there are systematic and structured factors\footnote{We refer to two closely related concepts of social and power structures. Power structure is discussed in~\ref{b:power}. Social structure is ``the patterns of relationships and distributions that characterize the organization of a social system. Relationships connect various elements of a system (such as social statuses) to one another and to the system itself. Distributions include valued resources and rewards, such as power and income, and the distribution of people among social statuses. Structure can also refer to relationships and distributions among systems''~\citep{Johnson2014-gj}.} that impede the effective operationalization of ethics in technical practice, including the following challenges (C1-C3):  
\begin{enumerate}[leftmargin=*,label=\textbf{C\arabic*}]
    \item \textbf{Abstract ethical principles vs. concrete realities.}
    Ethics principles and guidelines of conduct are typically formulated at a general and conceptual level.
    There is a lack of systematic support, incentives, and recognition to translate high-level ethical principles into low-level technical practice and ethical assessment~\citep{Bleher2023,Hagendorff2021,Munn2022,Ayling2021,Hagendorff2020,Morley2019,Mittelstadt2019,McLennan2022}. 
    \item 
    \textbf{Ethics for post-hoc fix vs. ethics encoded in the AI blueprint.}
    For those ethical principles that have been translated into technical practice, such as AI fairness techniques, ethics is not encoded from the outset in the blueprint of a technique~\citep{jin2025aiimagination,birhane2022values}, but is usually regarded as an add-on feature to fix ethical issues of the technique.  
    Ethics lacks the teeth~\citep{doi:10.1177/2053951720942541} to prevent problems from their roots and critically question, challenge, and prohibit techniques that are not aligned with ethical values. This phenomenon of prioritizing technical progress over ethics implies a systematic myth that frames ethics and regulations as a trade-off or impediment to research and technological innovations~\citep{schaakeTechCoupHow2024}. We argue against this myth. In fact, the above relationship of the promotion of rigorous science through ethics shows that ethics has a synergetic effect on technical innovation. Ethics plays the role of an opponent in an adversarial system, such as the discriminator in a GAN, the red team in software development, the brakes in a car, or the reviewers in peer review. The seemingly counteracting effect of ethics is necessary to prevent potential problems and ensure the overall healthy development of technologies. Therefore, like the brake to the engine, the strong momentum in technical advancement should be equipped with at least equivalently powerful ethics mechanisms for checks and balances. The current investigation of ethics in MISyn is far from achieving this goal. Therefore, we call for systematic changes in the research agenda to support, incentivize, and recognize more ethical works and practices in MISyn and, more generally, the MIA field. 

    \item \textbf{Ethics washing.} The implementation of ethics can bring about unethical issues, such as ethics washing that ``mak[es] unsubstantiated or misleading claims ... , or implementing superficial measures ... in order to appear more digitally ethical than one is''~\citep{Floridi2019}. Ethics washing happens when there are structural imbalances in power, and ethics is used as a means to serve the benefits and interests of the powerful, not as an end in itself for the interests and benefits of the most impacted people. This indicates that merely implementing and assessing ethics in a top-down manner is inadequate. Ethics should employ bottom-up approaches to ensure that stakeholders and the public, especially the marginalized, vulnerable, and mostly impacted, have the rights and power to shape the key aspects in planning, implementing, and assessing ethics in technical practice, such that their interests and benefits are reflected in technical development. 
\end{enumerate}

\begin{figure}[!ht]
    \centering
    \includegraphics[width=\linewidth]{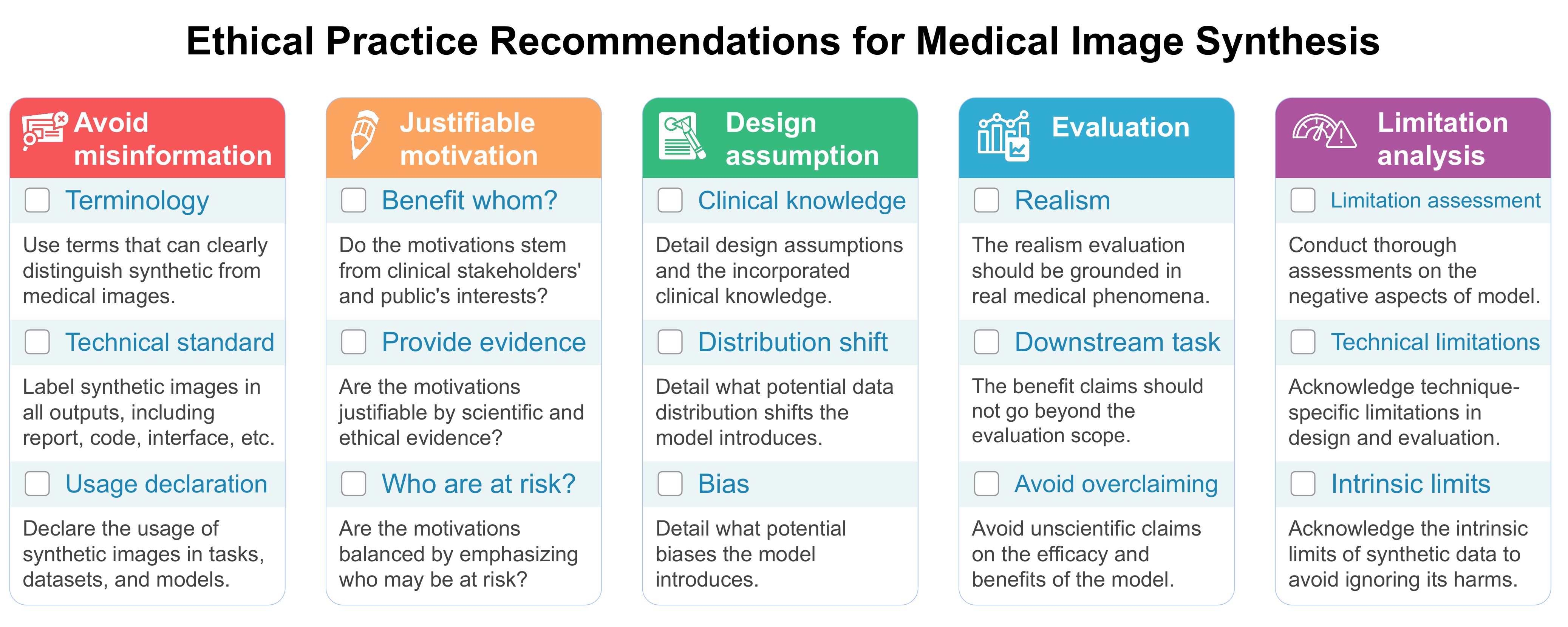}
    \caption{
    A checklist visualization of our proposed ethical practice recommendations for the technical development of MISyn in~\cref{action_points}. 
    \cref{app:case_study} demonstrates how to apply the recommendations in  paper review.
    }
    \label{fig:teaser}
\end{figure}

Our work aims to address the above challenges to facilitate the effective implementation of ethics in MISyn technical practice. Specifically, to address C1, our work is a demonstration of the process to encode ethical values and assumptions in the blueprint of technical design and evaluation. The process considers wide participation and oversight from stakeholders and the public, which avoids the ethics implementation being hijacked by unjust power and reflects the ideas in C3. 
Distinct from common technical ethics approaches that may reinforce questionable assumptions embedded in technology~\citep{greeneBetterNicerClearer2019a,birhane2022values, jin2025aiimagination}, 
we encode a different set of ethical assumptions that address C2 to allow ethics to play its adversarial role in promoting healthy development of technologies. Our work contributes to the ethical practice of MISyn by the following:

\begin{enumerate}[leftmargin=*]
    \item Based on ethical theories, assumptions, and principles, we perform an analysis of the general properties of ethical MISyn, and identify the epistemic limits and harms of MISyn.
    \item Based on the theoretical analysis, we provide practical support towards ethical MISyn, including ethical practice recommendations for technical practitioners (\cref{fig:teaser}), the oversight recommendations for non-technical stakeholders, and two case studies to demonstrate the process of applying the ethical practice recommendations. 
\end{enumerate}

Our paper has two parts: 
Part I is on ethical theories for MISyn (\cref{sec:assumption}-\ref{misinfomation}). We first set up the philosophical and ethical background assumptions in~\cref{sec:assumption}. Then we analyze the key properties of ethical MISyn in~\cref{define}, and identify the technical limits and ethical risks of MISyn in general and in the era of large pre-trained models in particular in~\cref{misinfomation}.

Part II is on  ethical practice for MISyn (\cref{action}-\ref{sec:case_study}). We propose a list of ethical practice recommendations to support the technical development of ethical MISyn in~\cref{action_points}, and a list of oversight recommendations to support non-technical statkeholders' oversight in~\cref{sec:oversee}. We conduct two case studies in~\cref{sec:case_study} to demonstrate how to apply the ethical practice recommendations in MISyn research, development, reporting, and review practive. 

To facilitate the association between the theoretical and practical parts, we link the suggested ethical practice recommendations (\ref{item:terminology} -- \ref{item:limit}) and oversight recommendations (\ref{cl:appropriate} -- \ref{cl:limit}) in~\cref{action} to their sources in Part I by referring to the item number in parentheses when discussing the rationale for a particular recommendation. 
\part{MISyn and Ethical Theories}

\section{The ethical and philosophical background assumptions}\label{sec:assumption}

Ethics is a branch of philosophy that involves normative judgment on which behaviors are morally right. As mentioned in the Introduction, since our approach is distinct from the common ethical approach in that it utilizes a different set of assumptions, in this section, we detail the ethical theories and assumptions we use. 
Ethical theories have different approaches shaped by various philosophical traditions to justify normative judgment ~\citep{Hagendorff2020,BIRHANE2021100205}. 
Different from the prevalent principled approach in AI ethics~\citep{Hagendorff2021} that is solely based on ethical principles (such as fairness, privacy, and accountability~\citep{ning2024generativeartificialintelligencehealthcare,10.1007/978-3-031-72787-0_17}), this work utilizes the following interrelated ethical approaches of critical theory~\citep{Waelen2022}, relational ethics~\citep{BIRHANE2021100205}, feminist ethics of care~\citep{sep-feminism-ethics,CostanzaChock2018}, power relations~\citep{Burrell2024,doi:10.1080/15265161.2024.2377139,BoenigLiptsin2022}, and epistemology~\citep{Fricker2007,CrasnowSharonL2024Feap} to augment the principled approach. We prioritize these ethical theories over the traditional theories used in AI ethics such as utilitarianism and deontology because:

\begin{enumerate}
    \item \textbf{More precise diagnosis.} The theories we choose use a critical perspective and encourage a grasp of ethical problems at their roots~\citep{BIRHANE2021100205}.

\item \textbf{More realistic assumptions.} These theories are based on more realistic background assumptions (see below) about technology, people, and society. In comparison, the traditional ethical theories of utilitarianism and deontology rely overly on their background assumptions about the ideal rational, abstracted, self-sufficient, and de-contextualized subject~\citep{BIRHANE2021100205,sep-feminism-ethics}.  

\item \textbf{Centering the less powerful for justice.} In the face of power asymmetry and marginalization, these theories seek ethical solutions that ``center the needs and welfare of those that are disproportionally impacted and not solutions that benefit the majority''~\citep{BIRHANE2021100205}. Solutions to protect the mostly impacted and vulnerable groups may enable us to better tackle the unethical issues at their roots, compared to solutions of doing good for the majority, because the latter cannot effectively prevent unethical issues, such as the medical scandals and algorithmic harms described in \cref{ethics}. The latter is also suspected of unethically framing the problem, as the capability to define what is ``good'' and who constitute ``the majority'' may be influenced by unjust power structures. 
Furthermore, prioritizing the interests of the less powerful over the majority also aligns with biomedical ethics that prioritizes patients' interests over the interests of science and society~\citep{rothman_strangers_1992}. 
 
\item \textbf{Ethics for better knowing.} We use the intersection between epistemology and ethics because the epistemological goal of pursuing true knowledge in MISyn activities aligns with the ethical goal of conducting just MISyn activities~\citep{Fricker2007,CrasnowSharonL2024Feap}.  
\end{enumerate}

\begin{figure}[!h]
    \centering
    \includegraphics[width=\linewidth]{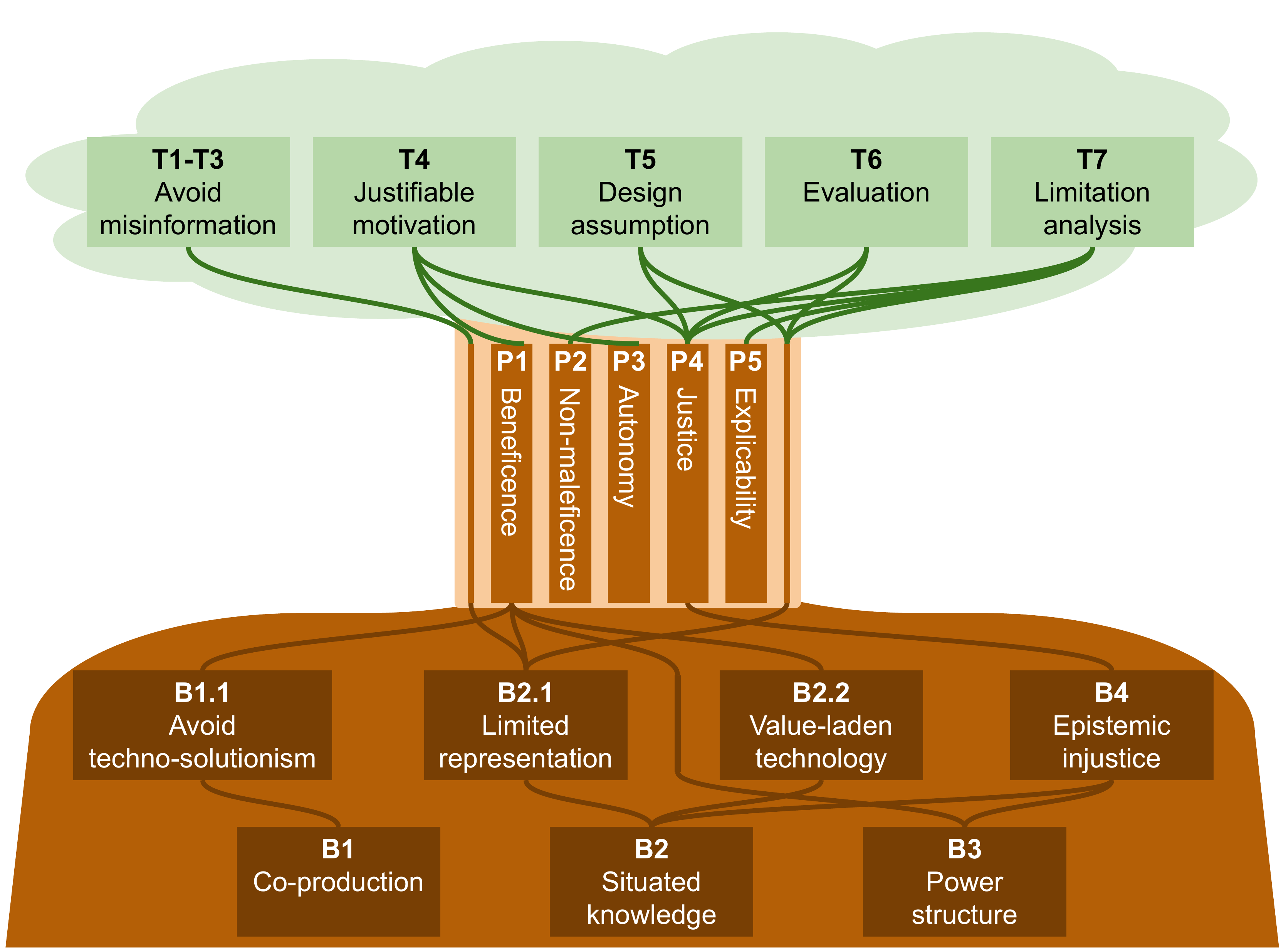}
    \caption{A tree structure visualizing the relationship of the background assumptions \ref{b:coproduce}-\ref{b: epistemic_injust} (roots) in~\cref{sec:assumption}, the ethical principles \ref{p:beneficence}-\ref{p:exp} (trunk) in~\cref{define}, and the ethical practice recommendations \ref{item:terminology}-\ref{item:limit} (leaves) in~\cref{action_points}. The roots and trunk of the ethical theories provide theoretical supports (visualized as lines) for the leaves of the ethical practice recommendations. 
}
    \label{fig:tree_bg_assmption}
\end{figure}

The background assumptions underpinning these ethical theories are different from the common background assumptions in the technical community~\citep{jin2025aiimagination}. Therefore, it is necessary to foreground these assumptions (denoted as B1, B1.1, B2, $\dots$ in the following paragraphs and in~\cref{fig:tree_bg_assmption}) upon which the subsequent ethical analyses of MISyn are based. 
Next, we list these background assumptions and provide their brief rationales. 
The detailed contents and supporting arguments of these assumptions and ethical theories are provided in \cref{app:bg} and in our prior work~\citep{jin2025aiimagination}. 

\FloatBarrier

\noindent 
 \textbf{Background assumptions (B):}
\begin{enumerate}[label=\textbf{B\arabic*}]

    \item \label{b:coproduce}\textbf{Co-production.} \textit{Technology and society co-produce each other. }
    
    Technology ``shapes and is shaped by the social world''~\citep{BoenigLiptsin2022}. Technology cannot be isolated and abstracted from the social conditions and relations that underpin it~\citep{BIRHANE2021100205,matthewman_technology_2011}. This is the sociotechnical lens to understand the role of technology~\citep{BoenigLiptsin2022}.
    \begin{enumerate}[label=\textbf{B\arabic{enumi}.\arabic*}]
        \item \label{b:soutionism}\textbf{Avoiding techno-solutionism.} \textit{Ethical issues may not be solved by technical solutions alone. }

        According to \ref{b:coproduce}, because ethical issues in technology are the co-production of technological and social factors, tackling the technological factor alone may not be able to solve problems at their roots.
    \end{enumerate}
\item \label{b:situated}\textbf{Situated knowledge.} \textit{Knowledge, including science and technology, has fundamental dependence on concrete lived experience.}

Because knowledge is grounded in and cannot be isolated from the concrete experience of the embodied and relational knowers who are socially located, knowledge always comes with a partial perspective of its knowers, which means knowledge is situated~\citep{CrasnowSharonL2024Feap,Haraway1988}. 

   \begin{enumerate}[label=\textbf{B\arabic{enumi}.\arabic*}]
   \item \label{b:nowhere}\textbf{Limited representation.} \textit{Knowledge, including science and technology, cannot fully represent the complex phenomenon of interest.}

   According to \ref{b:situated}, because all knowledge has a partial perspective, knowledge cannot grant us ``a window onto a disembodied, God’s-eye perspective''~\citep{frank_blind_2024} that can fully represent the complex phenomenon of interest.
    \item \label{b:value}\textbf{Value laden.} \textit{Knowledge, including science and technology, is value-laden.}

    Because knowledge is dependent on a particular perspective of its knowers (according to \ref{b:situated}), and knowers are socially located, the interests and values of knowers influence the process of knowing, and knowledge is neither value-free nor apolitical~\citep{CrasnowSharonL2024Feap,verbeekMoralizingTechnologyUnderstanding2011a,ChinYee2019}. 
    
    \end{enumerate}
    \item\label{b:power} \textbf{Power structure.} \textit{Power structures in a sociotechnical system ``lay out paths of least resistance that shape how people participate'' regardless of the individual participant's character 
    ~\citep{Johnson2014-gj}.}

    In a social system, power structure is the patterns and distributions of power, the ability to structure or alter the behavior of others~\citep{Johnson2014-gj,BoenigLiptsin2022}. Social system is larger than a collection of individuals because the social system also contains complex interactions and relationships as outlined by the social structure. This indicates that ``people are not systems and systems are not people,'' and it is possible that a system participated by good people can produce bad consequences because of the bad structure that sets paths of least resistance for good people to participate~\citep{Johnson2014-gj}.
        \item\label{b: epistemic_injust} \textbf{Epistemic injustice.} \textit{Epistemic injustice is the injustice ``when certain groups are systematically denied epistemic resources or are treated as non-epistemic agents (or less than fully human as knowers)''~\citep{CrasnowSharonL2024Feap}.}

        Epistemic injustice occurs because knowers are socially located in the social system (\ref{b:situated}),  and when the social system has asymmetric power structures (\ref{b:power}), the oppressed group of knowers can be wronged in their capacity as knowers due to their disadvantaged location in the power structure. Epistemic injustice is harmful to both the oppressed group and the knowledge community as a whole.
\end{enumerate}

With these background assumptions, in \cref{fig:tree_bg_assmption}, we visualize how these background assumptions and their ethical theories  support our ethical analyses of MISyn and ethical practice in subsequent Sections \ref{define}-\ref{action}.

\section{Properties of ethical MISyn}\label{define}

Ethical MISyn is defined as the task to ensure that the MISyn techniques are researched and developed ethically throughout 
their entire lifecycle:
from planning, development, and evaluation to deployment and maintenance. Regarding the criteria of being ethical, since ethics encompasses diverse values and theories, to determine the important ethical values in technical development, a democratic decision process with various stakeholders and citizens --- especially with the most impacted and marginalized groups to avoid epistemic injustice (\ref{b: epistemic_injust}) --- is needed. The democratic process is to ensure that the resulting techniques reflect public values and interests, as argued in~\cref{ethics}, and it is usually an iterative and non-linear process. As a primary step, in this work, we use the current AI ethics principles from~\citet{Floridi2018} as a proxy for the consensus from the democratic process. The AI ethics principles consist of five core values: beneficence, non-maleficence, autonomy, justice, and explicability~\citep{Floridi2018} (\cref{fig:tree_bg_assmption}). The first four are from bioethics that have been evolved and operationalized for decades to protect patients' and human subjects' rights, values, and interests. And the fifth one is newly added to tailor to AI technology.
Next, we apply the five AI ethics principles (\ref{p:beneficence}-\ref{p:exp}, the principles' content in bold text is used verbatim from~\citet{Floridi2018}) to the specific task of MISyn to delineate the key properties of ethical MISyn.

\begin{enumerate}[label=\textbf{P\arabic*},wide, labelwidth=!, labelindent=0pt]
\item\label{p:beneficence} \textbf{The beneficence principle: Promoting well-being, preserving dignity, and sustaining the planet.}

\ref{p:beneficence} encourages us to carefully inspect the purposes and objectives of MISyn to determine whether they are justifiable to align with the beneficence principle. The inspection can be conducted at the planning stage and throughout the lifecycle of MISyn~\citep{McLennan2022,BIRHANE2021100205}. 
Critically assessing the objectives of MISyn means to question the foundational premises. 
So instead of asking ``How to design an MISyn technique for good?'' one could examine the premises of this question, such as, ``Is the problem formulation justifiable and appropriate to be solved by an MISyn technique?'' 
and ``What does beneficence mean, and for whom?'' We unpack the two examination points below. 

$\bullet$ \textbf{Questioning the appropriateness of MISyn.} 
Critically examining the appropriateness of MISyn is essential because not every problem is suitable to be solved by MISyn techniques (according to the background assumption \ref{b:soutionism} of avoiding techno-solutionism): examination prevents us from framing MISyn as the default option in problem solving. Such a framing effect may impede us from understanding the root causes and seeing other possibilities in problem solving. The framing effect may also tend to make people downplay the negative aspects of MISyn, such as its intrinsic limits, risks (as we will analyze in~\cref{misinfomation}), and costs (such as the significant energy, water, and other natural resource consumptions involved in training and maintaining MISyn models~\citep{JMLR:v21:20-312,li2023makingaithirstyuncovering,crawfordAtlasAIPower2021}). 

To examine the appropriateness of MISyn, one could examine whether MISyn genuinely promotes the beneficence principle, particularly in comparison to alternative approaches, such as no action (e.g., questioning whether it is really a problem that needs addressing, or it is a fake problem merely driven by the values and interests of the technical sector), non-technical, low-technical, or other technical solutions (\ref{item:appropriate}, \ref{cl:appropriate}).
For instance, considering one of the main motivations of MISyn is to augment data and mitigate data scarcity, depending on the specific application context, a non-technical or low-technical alternative for this motivation could be to enhance the workflow of clinical data collection, consensual usage, and privacy protection to increase the availability of medical images;
a technical alternative could be to utilize federated learning techniques~\citep{10.1016/j.patcog.2024.110424} that may potentially increase the availability of medical images while protecting patient privacy.
Both alternatives address the disadvantages and inherent limits of MISyn, while also increase the number of available medical images that are grounded in real medical phenomena. 
The examination should be embedded in social context to incorporate multiple stakeholders' --- not merely the technical sector's --- values, inputs, and perspectives, as detailed in the next point. 

$\bullet$ \textbf{Who benefits and who is harmed?}
Asking what is ``good'' or ``beneficence'' is a value judgment. Our background assumptions reject a universal and de-contextualized definition of good. Values are a complex and irreducible phenomenon. According to the background assumption of \ref{b:nowhere} limited representation, there is no such a thing as a ``universal'' value that can fully represent the pluralistic and contextualized nature of values for being good. A ``universal'' good for the majority is an oversimplification of the irreducible phenomenon of values, which can cause oppression to the marginalized groups who hold diverse values. Furthermore, according to \ref{b:nowhere}, any abstraction of the values of being good reflects  certain partial perspective of its knower, and there is no such a thing as a ``neutral'' good that is detached from perspectives of its knower. The seemingly perspective-free good for the majority can easily be used to disguise values and perspectives of the dominant power, which could be unjust. Therefore, to avoid the above unethical pitfalls of the definition of a universal and neutral beneficence for the majority, it is more preferable to adopt a pluralistic perspective of beneficence that beneficence is always associated with the perspective of its knower. This means that the goal of doing good should always specify ``for whom?''

Putting this idea into the practice of MISyn means that setting the usual objectives in the technical community as the goals for MISyn --- goals that usually legitimize themselves by aiming to maximize the beneficence for the majority --- may not be ethically adequate for healthcare. A recent empirical study with medical, ethical, legal, and technical experts who work with medical AI shows that the technical objectives in medical AI do not align with medical and ethical needs in the healthcare domain~\citep{ArbelaezOssa2024}. 
Based on the above analysis, because ``a universal and natural good for the majority'' is an oversimplification of values and obscures the perspective of its knowers, setting it as the technical objective for MISyn can easily fall into the above unethical pitfalls of: 1) oppressing the values, benefits, and perspectives of the marginalized in the name of pursuing universal good, and 2) serving the perspectives, interests, and benefits of the powerful in the disguise of pursuing a perspective-free good for the majority. 
The unethical issues are more concerning right now, because the current dominant power structures in the technical community have a tendency to concentrate power and tend to encode values and perspectives of the private sector --- rather than the public or society --- when pursuing technical objectives such as novelty, performance, efficiency, and generalization~\citep{birhane2022values,10.1145/3531146.3533194,Burrell2024}. 
Thus, instead of using the common technical objectives to  do ``good'' for the ``majority'' as the ethical justification for MISyn in healthcare, a more ethical practice is to acknowledge the limited perspective associated with value judgments, and critically justify why and how an MISyn technique is appropriate to benefit the target group, and which groups could be harmed by the technique whose values and benefits are out of the focus of the limited perspective. 
To make an MISyn technique beneficial to the target healthcare stakeholders, we should explicitly align the technical perspective with the healthcare stakeholders' perspective, and incorporate the healthcare stakeholders' values, interests, and benefits in MISyn technical development (\ref{item:appropriate}, \ref{cl:appropriate}).

\item\label{p:non-malefience} \textbf{The non-maleficence principle: Privacy, security and ``capability caution.''}

The use of MISyn techniques should avoid potential harms, including harms caused by inappropriate use of patients' data (violating patients' privacy~\citep{Kaabachi2025}, the consent process, and data security), and harms caused by not understanding the ``upper limits on future AI capabilities'' and misuse~\citep{Floridi2018} (\ref{item:limit}, \ref{cl:limit}).
In~\cref{misinfomation}, we will analyze the limits and ethical risks of MISyn models, especially the risk of misinformation.

\item\label{p:autonomy} \textbf{The autonomy principle: The power to decide (whether to decide)}.

The autonomy principle suggests that the use of MISyn models and their downstream clinical decision support systems (CDSS) should support, rather than suppress, clinical users' and patients' autonomy in clinical decision-making. Autonomy also indicates that clinical users and patients have the right to refuse the deployment of algorithms, the right to refuse decisions from algorithms, and the right to know the correct information, including the limitations and weaknesses of technologies to avoid being misled by overclaims and overstatements of technologies~\citep{narayanan2024ai}.
This is related to the misinformation risk of MISyn that will be discussed in~\cref{sec:risk}. 
To achieve this goal, it is crucial for clinical users and patients to have the power to decide
in the planning and deployment of MISyn models, CDSS, and other clinical systems,
and avoid power asymmetry between technical people and clinical stakeholders (\ref{item:appropriate}, \ref{cl:appropriate}). Setting up power check mechanisms indicates that ``the burden of evidence for justifying why the deployment of predictive optimization is not harmful should rest with the developers of the tools,'' as~\citet{10.1145/3636509} suggested.

\item\label{p:justice} \textbf{The justice principle: Promoting prosperity and preserving solidarity}.

The justice principle is to ensure that the benefits and prosperity of MISyn techniques are shared and distributed fairly, and the potential harms are not disproportionately borne by marginalized social groups. As we will analyze in~\cref{misinfomation}, a limit of MISyn models is their potential to introduce additional biases in synthetic data, which could negatively impact patients from
marginalized communities in real-world applications.
This risk should be fully disclosed and properly addressed in the design and evaluation of MISyn techniques (\ref{item:design}, \ref{item:limit}, \ref{cl:limit}).

\item \label{p:exp} \textbf{The explicability principle: Enabling the other principles through intelligibility and accountability}.

The transparency and accountability principle guides the implementation of the aforementioned ethics principles. The development of MISyn models should be transparent and disclose the details of data collection, data source, training, and evaluation methods. Since the target application domain of MISyn is healthcare, the current MISyn reporting standards from the technical community are not enough because, ideally, the reporting should follow higher standards that comply with clinical research and development criteria, in order to enable stakeholders to have a full understanding of the capabilities of the MISyn techniques~\citep{HernandezBoussard2020}. For example, the report should include a thorough analysis of the scope, limitations, weaknesses, and failure cases of the MISyn technique (\ref{item:limit}). The evaluation metrics should be carefully selected to tailor to the specific clinical problem and data~\citep{MaierHein2024}.
\citet{Burnell2023} suggested that the aggregated metrics, such as overall performance on the test set, are insufficient to reflect ``important features of the problem space.''
They suggested that the report should include performance breakdowns across features in the problem space, and ``instance-by-instance evaluation results'' to enable granular and subgroup analyses.
The MIA community should also set up review standards and guidelines to disclose conflicts of interest and funding, as in biomedical research~\citep{10.1145/3531146.3533194,birhane2022values,stengaard2020} (\ref{cl:limit}).
In addition, developers of MISyn techniques should be accountable to correct mistakes and harms caused by MISyn. 
\end{enumerate}

Many of the above ethical MISyn properties are shared with the ethical requirements for other AI-related MIA techniques. In the next section, we provide a thorough analysis of the technical limits and risks that are specific to MISyn techniques.

\section{Theoretical analysis of the limits and risks of MISyn}
\label{misinfomation}

In this section, we conduct a theoretical analysis of the epistemic limits and ethical risks of MISyn. First, we set up the problem to analyze the epistemic grounding of MISyn in~\cref{setup}. Next, we deduce the intrinsic epistemic limit of MISyn in~\cref{sec:risk}; we show that, unlike real medical images, MISyn samples do not automatically establish grounding in real medical phenomena; 
not acknowledging this limit permits the misinformation of MISyn. 
Further in~\cref{harm}, we analyze the direct and indirect harms of misinformation; we reveal another epistemic limit of MISyn that synthetic data cannot fully resemble their training data in distribution and biases; we also examine the shortcomings of existing evaluation methods, which fail to recognize the epistemic limits and harms of MISyn. 
Lastly, in~\cref{llm}, we analyze the ethical risks of a particular type of MISyn: the foundation model-based MISyn.

\begin{table}[h]
    \centering
    \begin{tabular}{p{3cm}|p{12cm}}
    \toprule
       \textbf{Abbreviation or symbol}  & \textbf{Description} \\
       \hline
       AI & Artificial Intelligence \\
       CDSS & Clinical decision support systems\\
   CV & Computer Vision \\          
   DS3D & DermSynth3D \\
       FM & Foundation Model \\
       K-D model & Knowledge-dataset model \\
 MISyn  &  Medical image synthesis \\
 RCT & Randomized controlled trial \\
       REB & Research ethics board  \\
       \hline \hline
       $\Gamma$ & Real-world medical phenomena \\
       $\Omega$ & A model  \\
       $\Omega_1$ & Knowledge-dataset model \\
        $\Omega_2$ & Medical image synthesis model \\
       $\Omega_3$ & Downstream model \\
       $x^i$ & An instance drawn from a model space, with $i\in \mathbb{N}$ denoting the order of instance in natural number\\
       $\Longrightarrow$ & Grounding authenticity\\
       $\longrightarrow$ & Symbolic authenticity\\
\bottomrule
    \end{tabular}
    \caption{List of abbreviations and symbols used in the paper.}
    \label{tab:my_label}
\end{table}

\subsection{Problem setup: The epistemic grounding of MISyn}\label{setup}

We analyze the epistemic limit of MISyn by first constructing its epistemic framework. The epistemic grounding of MISyn refers to the knowledge sources that the MISyn techniques are based on, i.e., 
where the MISyn models acquire their knowledge from.
We first provide a general axiom of the relationship between a model and its modeling phenomena, and then analyze the epistemic grounding of MISyn from this first principle of \cref{axiom1}.

\begin{axiom}    \label{axiom1}
    Any model or combination of models is only an approximation, but not the full representation, of the complex phenomenon it aims to model.
\end{axiom}
We define a model as an entity that contains the extracted information, theory, rule, knowledge, insight, etc. from a phenomenon. In the context of MISyn, 
models can be in the form of
medical datasets, medical knowledge, clinical experience, and computational models.
We define the complex phenomenon using the same definition of complex systems, which are ``co-evolving multilayer networks,'' are context-dependent, and are composed of many non-linearly interacting elements~\citep{10.1093/oso/9780198821939.003.0001}. In the context of MISyn, the complex phenomenon can be the real-world medical phenomenon, medical knowledge, and medical experience. 

\cref{axiom1} is backed up by a common aphorism in statistics that ``all models are wrong, but some are useful''~\citep{Box1976}, and similar ideas on scientific models and theories have been discussed in the background assumption \ref{b:nowhere} limited representation, ~\cref{worldview} and~\cref{tab:blindspot} of the worldviews of science and technology. 
\cref{axiom1} emphasizes that models can be regarded as useful abstractions and simplifications --- bounded by their scopes, assumptions, and constraints --- to approximate certain aspects of the phenomenon, but cannot be regarded as the phenomenon itself because the phenomenon is not bounded~\citep{Cartwright_1999,10.1162/artl_a_00336}.
Therefore, we can neither claim that models can have a full representation of the phenomenon, nor use models to replace the phenomenon without fully acknowledging and disclosing the limits of such a replacement~\citep{Birhane2022}, i.e. ``surreptitious substitution''~\citep{frank_blind_2024} (\ref{item:limit}, \ref{cl:limit}).

\begin{figure}[!h]
    \centering
    \includegraphics[width=0.85\textwidth]{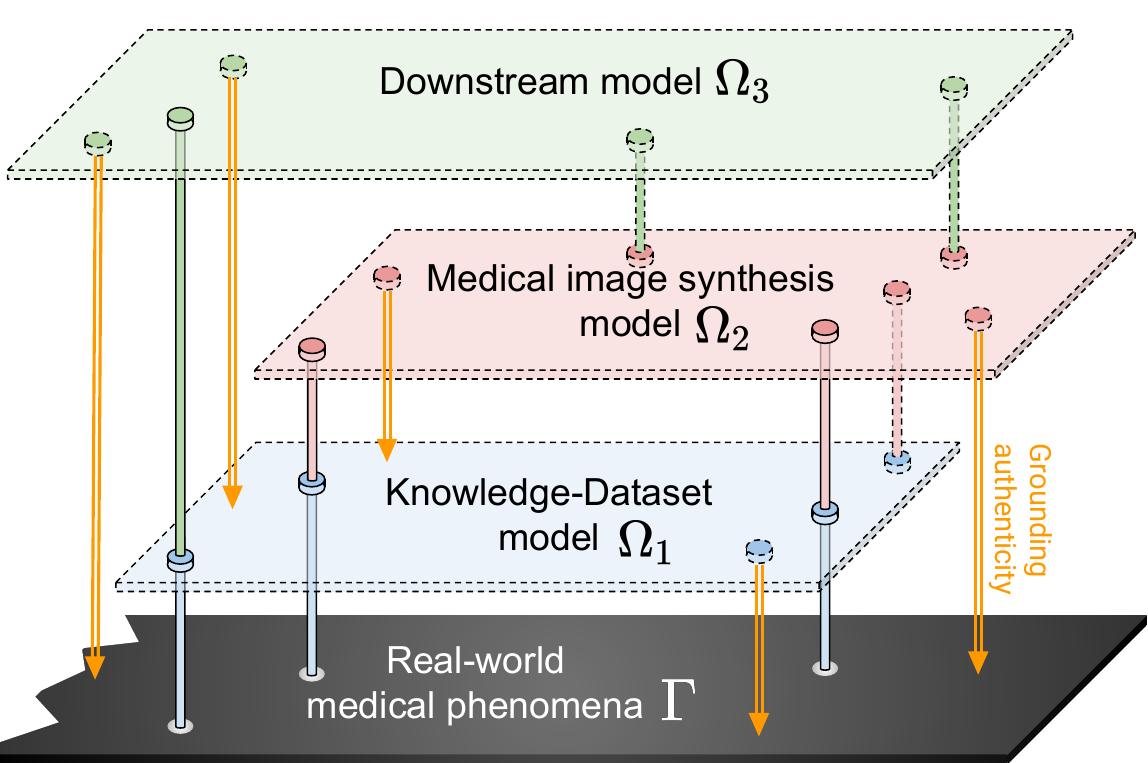}
    \caption{
Visualization of the relationship among the real-world medical phenomenon $\Gamma$, an MISyn model $\Omega_2$, its knowledge source $\Omega_1$, and its downstream model $\Omega_3$. $\Gamma$ is visualized as the solid ground: it is solid because $\Gamma$ is the reality; it is visualized as the ground because $\Gamma$ provides the epistemic grounding for the floors built upon it.   $\Omega_{1-3}$ are visualized as floors with dashed outlines. Floor indicates that $\Omega_{1-3}$ are imaginary spaces, and dashed outline indicates their abstract nature as opposed to the concrete reality. The solid dots and their corresponding pillars supporting a floor indicate specific instances/observations to support the model. A floor can be regarded as the imaginary space spanned by dots that has established grounding in the target space (has a pillar with the same color as the floor). The dashed dots are new hypothetical instances deduced within the model and exhibit symbolic authenticity. The dashed dots can also support an upper floor, visualized as the pillar. The newly inferred dashed dots (dots that do not have a pillar beneath them) do not automatically ground in a lower space, unless we explicitly establish grounding authenticity (visualized as the downward arrows) for them. 
    }
    \label{fig:grounding}
\end{figure}

Based on \cref{axiom1}, we can analyze the epistemic grounding of MISyn, which is the relationship between a model and the phenomena it models.
A MISyn model is a computational model with two knowledge sources: (i) its training dataset, composed of real medical images, and (ii) human knowledge, encoded as model priors, such as model architectures, optimization choices, and loss functions.
We denote (i) and (ii) combined
as the Knowledge-Dataset (K-D) model, and it forms the epistemic grounding of the MISyn model\footnote{Note that for cases where no human knowledge is explicitly used in the MISyn modeling process, the MISyn model is still grounded in the knowledge-dataset model.
This is because human knowledge is always implicitly encoded in the data collection process (such as the diagnosis or inclusion and exclusion criteria of medical images in the dataset), and there is no theory-free dataset~\citep{ChinYee2019,Andrews2024} (also see \ref{b:value}).
}, as illustrated in~\cref{fig:grounding}. 
We use $\{x_2^i|x_2^i\in \Omega_2, i\in \mathbb{N} \} \Longrightarrow \Omega_1$ to denote that an instance $x_2^i$ in the MISyn model $\Omega_2$ has established its grounding in the K-D model $\Omega_1$ (referred to as grounding authenticity as introduced below), also illustrated as pillars in~\cref{fig:grounding}. 
The grounding authenticity is established by the alignment evaluation of how the generated images resemble real images, i.e., the realism evaluation of synthetic images~\citep{kynkaanniemi2019improved}. 
The K-D model also has its epistemic grounding: the medical knowledge and datasets are obtained from the real-world medical phenomena $\Gamma$ via medical practice, observations, and interactions with patients.
This established grounding is denoted as $\{x_1^i|x_1^i\in \Omega_1, i\in \mathbb{N} \} \Longrightarrow \Gamma$,
where $x_1^i$ is a specific observation or a sample drawn from the K-D model $\Omega_1$, and $i$ is the instance ID. 

\sloppy To ensure that we can truthfully infer information within and across spaces of $\{\Gamma, \Omega_1, \Omega_2, \Omega_3, \dots , \Omega_n \}$,\footnote{Since a model $\Omega$ can be built upon previous models, there can be an infinite number of $\Omega$s.}
we define two types of authenticity\footnote{The two types of authenticity are from Jin Guantao's philosophy of authenticity~\citep{guantao_real_2023} and are grounded in the philosophy of science regarding the expansion of knowledge through scientific methods
\citep{frank_blind_2024}.}: symbolic authenticity ($\longrightarrow$) and grounding authenticity ($\Longrightarrow$). 
Symbolic authenticity refers to reasonable inference logically or mathematically within a model, 
such as model inference within its scope and assumptions. 
Grounding authenticity means that an instance in a model has established grounding in a specific lower space via scientific evidence,
indicated by the orange top-down vertical arrows from a source to a target space in~\cref{fig:grounding}. 
We distinguish the two types of authenticity because one cannot be confused with the other: symbolic authenticity is to establish grounding within a model, while grounding authenticity is to establish grounding across spaces.
Furthermore, symbolic authenticity is established via symbolic manipulation (logically or mathematically) on instance or space level, whereas grounding authenticity can only be established via empirical evidence on specific instances. 
As illustrated in~\cref{worldview} and~\cref{fig:worldview}-A, the combination of symbolic and grounding authenticity enables us to 1) establish scientific theories and computational models from observations and instances (grounding and symbolic authenticity), 2) infer (symbolic authenticity), and 3) test (grounding authenticity) new scientific hypotheses and computational models to expand our knowledge and understanding about the real world. 

\cref{axiom1} reveals the intrinsic epistemic limit of models.
Not acknowledging such limits and ``surreptitiously substituting''~\citep{frank_blind_2024} computational models for reality puts us at risk of misinformation of MISyn, as detailed in~\cref{sec:risk}. 

\subsection{The misinformation risk of MISyn}\label{sec:risk}

\begin{definition}[Misinformation of MISyn]
We define misinformation of MISyn as:
\begin{enumerate}[label=(\roman*)]
    \item providing or presenting synthetic images in ways or in contexts that may lead people to assume them to be real medical images, or
    \item providing or presenting computational systems that are fully or partially built on synthetic images in ways or in contexts that may lead people to assume them to be fully built on real medical images. 
\end{enumerate}
\end{definition}

In either case, the source of misinformation is the behavior of substituting synthetic images for medical images without acknowledgment (\ref{item:terminology}, \ref{cl:misinformation}). 
Misinformation is usually caused by fault or negligence in the technical development that mistakes the epistemic grounding of synthetic images (which will be shown in~\cref{theorem1} below), and ignores the following grounding assumption in the medical community\footnote{Wrongly ignoring the common assumptions in the medical community could be a manifestation of epistemic injustice in technical development, as discussed in~\cref{epistemic_injustice}.}: In the medical community, people always assume that medical images are acquired from patients and are grounded in real-world medical phenomena;
the notion of images being generated \textit{in silico} is not typically considered, as generative images in clinical settings are relatively new, and the medical community does not traditionally deal with them. Although there are creations such as medical illustrations and 3D human models for surgical simulations, these creations --- if indistinguishable from real medical images at a glance --- are labeled as artistic creations or simulations to distinguish them from real medical images. 

From the comparison between medical illustrations and medical images, we want to emphasize that the source of misinformation is not the behavior of generating images that imitate real medical phenomena.
As an analogy, we do not see fictitious stories, such as fictions and movies, as a problem compared to non-fictions and documentaries, because they are declared to be fictional from the outset. In contrast, fake news is a problem because it takes advantage of people's assumption about \textit{news}, which assumes the information has factual grounding in the real world. 
Similarly, misinformation of MISyn can happen when taking advantage of people's assumption about the notion of \textit{medical images}, which are assumed to be from patients. 
This is called ``violation[s] of representational assumptions''~\citep{Offenhuber2024} (\ref{item:standard}, \ref{cl:terminology}).

Since misinformation is rooted in tacitly substituting synthetic images for medical images, and thus tacitly using real-world medical phenomena to replace the original grounding of synthetic images\footnote{As analyzed in~\cref{setup}, synthetic images are grounded in their MISyn model (if they fulfill symbolic authenticity) or the K-D model (if they fulfill both symbolic authenticity and grounding authenticity in the K-D model).} (“the surreptitious substitution” in~\citet{frank_blind_2024}'s term),
we now ask: 
is it ever possible to legitimately replace the grounding of synthetic images and claim that the synthetic images are grounded in real medical phenomena, provided that the MISyn model has good alignment with the K-D model, \ie, can generate highly realistic images? The short answer is no, because the real-world medical phenomena are irreducible to the MISyn or K-D models according to~\cref{axiom1}. We put the rationale formally in the proof of~\cref{theorem1}.

\begin{theorem}
    For a medical image synthesis model, being realistic (i.e., having good alignment with the training datasets and medical knowledge) 
    does not ground its synthetic data in real medical phenomena.
    \label{theorem1}
\end{theorem}

\begin{proof}
    A medical image synthesis (MISyn) model 
    approximates the distribution of the training inputs, including the real medical image datasets, and/or medical knowledge, which we name the K-D (knowledge-dataset) model. According to \cref{axiom1}, the MISyn model can be an approximation of the K-D model, but does not have the full representation of the K-D model. The realistic metrics measure the degree of approximation to indicate how well the MISyn samples are grounded in or aligned with the K-D model. Then, 
    according to \cref{axiom1},     
    there will always be MISyn samples that cannot be grounded in the K-D model to establish their grounding authenticity.
    
    For those MISyn samples $\{x_2^i |x_2^i\in \Omega_2, i\in \mathbb{N} \}$ that can establish grounding authenticity in the K-D model $\Omega_1$, 
    denoted as $\{x_2^i|x_2^i\in \Omega_2, i\in \mathbb{N} \} \Longrightarrow \Omega_1$ where $\Omega_2$ is the MISyn model, these samples can be regarded as reasonable interpolation within the K-D model, thus fulfilling the definition of symbolic authenticity, \ie, $ \Omega_1 \longrightarrow \{x_2^i|x_2^i\in \Omega_2, i\in \mathbb{N} \} $.
    Despite this, we cannot confuse symbolic authenticity with grounding authenticity. 
    According to \cref{axiom1}, the K-D model is only an approximation of, and cannot fully represent, the real medical phenomena. Thus, the K-D model does not automatically establish grounding authenticity in real medical phenomena. The way for instances sampled from the K-D model to establish grounding authenticity is to identify patients who present the same imaging phenomena by empirical evidence. Therefore, these synthetic samples must find their real-world counterparts before we can further establish additional grounding authenticity to claim that they are grounded in real medical phenomena. 
    Since we cannot establish the grounding of the MISyn samples unless we find real-world evidence from patients, we cannot claim that the MISyn samples are grounded in real medical phenomena, despite some of them having highly realistic measures. 
\end{proof}

From~\cref{theorem1}, it is straightforward to get the following~\cref{corollary1}:
\begin{corollary}
    For medical image synthesis models, being realistic (i.e., having a good alignment with the training datasets and medical knowledge) does not preclude misinformation risk.
    \label{corollary1}
\end{corollary}

\begin{proof}
    According to~\cref{theorem1}, the fact that synthetic images are highly realistic does not provide empirical evidence to ground them in real medical phenomena. Therefore, mistakenly claiming that such synthetic images (regardless of the realism) being grounded in real medical phenomena surreptitiously substitutes their grounding, satisfies the definition of misinformation of MISyn. 
\end{proof}

Theorem~\ref{theorem1} establishes the fact that the synthetic images do not automatically have groundings in real medical phenomena, unless being empirically proved by identifying patients who exhibit the phenomena in synthetic images, which could be very costly and almost unrealistic to acquire\footnote{Finding patients who exhibit the same phenomena of the $n$ synthetic images is unrealistic and almost impossible, because it is way more costly than acquiring $n$ number of medical images: the cost would be the sum of the costs to acquire $m$ ($m \gg n$) medical images, and to search, verify, and match the $n$ synthetic images to the $m$ medical images.}. 
This is the intrinsic epistemic limit of synthetic images that do not automatically establish grounding in real medical phenomena\footnote{
Due to the intrinsic limit in the epistemic grounding of synthetic images, it is impossible for synthetic images to have the ``gold standard'' ground truth labels for diagnostic or classification tasks, even if they are synthesized based on certain diagnoses. This is because the gold standard of medical diagnostic labels is usually acquired from the synergized information of patients' clinical, radiological, and pathological diagnoses. Synthetic images have, at best, the silver standard ground truth label from the consensus of medical experts' interpretation.}, unlike real medical images. Failure to see the epistemic limit permits the misinformation of MISyn.

\subsection{The epistemic harms and benefits of MISyn}\label{harm}

\subsubsection{The direct and indirect epistemic harms of MISyn}
The misinformation risk comes from misattributing the grounding of MISyn samples. Now, we analyze the direct and indirect ways in which misinformation of MISyn is unethical and detrimental to the medical community. 

The direct harms of misinformation of MISyn to the medical community are similar to the harms of fake news to the public. 
If the synthetic images are not explicitly labeled as synthetic, people in the medical community can assume that the indistinguishable ``medical images'' came from patients. This can mislead healthcare professionals' learning to mistake the potentially nonexistent phenomena exhibited in synthetic images as real medical phenomena. Misinformation pollutes the medical image data environment, because users will need to put in extra effort to verify whether these images are synthetic or real. Furthermore, even if no misinformation is present, the existence of the misinformation phenomenon endangers credibility and erodes trust in medical images. Because users may raise doubt on the authenticity of images even if they are from patients, and such doubts were unnecessary in the pre-misinformation era. 

The indirect and far-reaching harms of misinformation are when the synthetic data are used for downstream models, datasets, and tasks. If the downstream models do not explicitly disclose that the models were trained fully or partially on synthetic images, model users will reasonably presume that real medical images were used during training. This is misinformation of MISyn for downstream model users (\ref{item:declare}, \ref{cl:misinformation}).

In the medical community, the source of data is an important piece of information for healthcare professionals to interpret conclusions drawn from data. 
For example, when interpreting the result of a clinical trial, physicians need to check the trial population, and the inclusion and exclusion criteria of patients to see how well the conclusion from the trial can be applied to their patients~\citep{Sonbol2020,Zarbin2017}. 
This is the same with AI. In a user study with neurosurgeons on the clinical utility of AI-assisted decision support, a doctor mentioned training data as a source to establish credibility of the AI system ``to make sure that I’m inputting [data with] the same [distribution], that my [MRI] scan is no different than [the training dataset]''~\citep{JIN2024102751}.

The doctor's comment underlines the importance of disclosing the use of synthetic images in downstream tasks, datasets, and models, because synthetic images cannot be regarded as the full representation of real medical images based on \cref{axiom1}, no matter how well they approximate the real medical images. Synthetic images differ from medical images in expected and unexpected ways with respect to the assumptions of acquiring/generating the data, the data distributions, implicit and explicit biases in the data, etc. (\ref{item:design}). This poses new challenges for how we evaluate and report MISyn techniques, as detailed in the next subsection.

\subsubsection{Rethinking MISyn evaluation}\label{sec:eval}

Currently, there are two general approaches to MISyn evaluation~\citep{DAYARATHNA2024103046} regarding establishing a valid grounding in the K-D model and the practical utility: (i) one focuses on the quality of the synthetic images by assessing their realism, and 
(ii) the other assesses the epistemic benefits of the synthetic data for downstream task models, 
such as performance improvements in classification, segmentation, or registration task.
However, the latter evaluation is lopsided, as it only evaluates the epistemic benefits while ignoring the epistemic harms of MISyn techniques in downstream tasks in the long run and in the real world. Next, we use epistemic grounding analysis to show why such performance evaluations alone cannot address the epistemic harms of MISyn. 

Task performance usually measures how well the downstream model $\Omega_3$ grounds its learning in the training data including synthetic data, which confines the assessment of the grounding authenticity to the joint spaces defined by MISyn model $\Omega_2$ and K-D model $\Omega_1$ (\Cref{fig:grounding}), denoted as $\{x_3^i|x_3^i\in \Omega_3, i\in \mathbb{N} \} \Longrightarrow \{\Omega_1,\Omega_2\}$ where $x_3^i$ is an instance drawn from the downstream model $\Omega_3$. 
This evaluation paradigm lacks a thorough assessment on external validation dataset of well-designed benchmarks~\citep{Burnell2023} on how well the downstream models $\Omega_3$ ground their learning in real medical phenomena $\Gamma$, $\{x_3^i|x_3^i\in \Omega_3, i\in \mathbb{N} \} \overset{\text{?}} \Longrightarrow
\Gamma$.
Because the \textbf{far-reaching epistemic harms of synthetic images happen in the real medical phenomena $\Gamma$} when the downstream models are used in medical settings, the performance evaluation --- confined within the scope of the model space $\{\Omega_1,\Omega_2\}$ --- is not enough to assess the epistemic harms of MISyn in the real world $\Gamma$. 
This reflects the inherent gap between the epistemic limits and their real-world impacts, as detailed below.

Regarding the epistemic limits, the MISyn modeling process inevitably introduces additional distribution shift and biases, compared to the original distribution and biases of the data it was trained on.
A recent study provides empirical and theoretical evidence to show that it is ubiquitous and inevitable for data-driven generative models and their downstream models to exhibit the central tendency of the data distribution (tails disappear) and smoothen the data distribution (reduced variance)~\citep{Shumailov2024}, thereby erasing diversity in the real medical phenomena; the authors term this degenerative learning phenomenon ``model collapse.'' 
Even for meticulously designed techniques 
to explicitly augment the long tail phenomena (to increase diversity in data distribution) and/or to reduce certain biases in the original dataset, it is impossible for such approaches to capture all the real-world diversity or exclude any biases
(otherwise, it will violate \cref{axiom1}). 

Regarding their real-world impacts, when the MISyn-trained downstream models are used in clinical settings, patients whose disease manifestations are smoothed out by the synthetic dataset could suffer from poor performance of the downstream models, and marginalized groups may suffer from the new unknown biases introduced by MISyn models.
Without rigorous monitoring and quality control of the CDSS, these harms may not be readily identified and corrected, because inflicted people usually lack the resources to realize that they are being wrongly treated in the health system~\citep{Carel2014}. 
In other words, it is the powerful who enjoy the benefits of MISyn (successfully persuading doctors to accept the so-called powerful AI technologies), while the harms are disproportionately distributed to the powerless. This violates the ethical principles of justice and non-maleficence, and is a manifestation of epistemic injustice and its silenced harms as discussed in the background assumption~\ref{b: epistemic_injust} and~\cref{epistemic_injustice}.
Rather than adapting the models' learning capabilities to the diversity of real medical phenomena, we may adapt the real-world tasks to the capacity of downstream models using synthetic data, creating an illusion of progress in performance within spaces defined by models. 
``We may think we're making artificial systems with their own genuine autonomy, but we’re actually remaking our environment to fit the limitations of our non-autonomous devices,'' as remarked by~\citet{frank_blind_2024}. 
Using the model performance improvement within the model space as the end point of MISyn assessment to justify their real-world benefits is ``a case of surreptitious substitution, of substituting computational states for the imprecise and fluid everyday world''~\citep{frank_blind_2024}, as we analyze in~\cref{worldview} and~\cref{tab:blindspot}. 

The analysis indicates that the existing evaluation of performance improvement in downstream tasks does not touch on the epistemic harms of MISyn techniques. 
To improve the scientific rigor and ethics of the technical evaluation and reporting of MISyn, we propose the following three directions based on our analysis:
\begin{enumerate}[leftmargin=*]
    \item \textbf{Limitation analysis. } In the medical community, it is unacceptable to only report the benefits of a medication or medical device without thoroughly investigating its scopes and side effects. 
    In contrast, the current technical community generally lacks awareness of including rigorous and thorough limitation analysis in technical evaluation and reporting standards, and we associate it with epistemic injustice in the background assumption~\ref{b: epistemic_injust} and~\cref{epistemic_injustice}. 
    Therefore, the first thing we propose to change is to include the reporting of epistemic harms and limitations in MISyn evaluation (\ref{item:limit}, \ref{cl:limit}). The evaluation should investigate questions such as: What are the effects of introducing synthetic data on the overall data distribution, especially on the long-tail distribution of image features and disease classes? What potential biases are introduced in the synthetic dataset? Since it is not feasible to identify all possible deviations of the synthetic data from the real data, we should always declare this fact as one of the limitations of MISyn techniques.
    
\item \textbf{Real-world grounded evaluation. } 
We propose to investigate and implement methods and standards to ground the assessment of MISyn, including its benefits and harms, in real-world medical phenomena. We detail this consideration below regarding the two evaluation aspects on realism and downstream task performance. 

Regarding the evaluation on realism, 
because MISyn models are to establish epistemic grounding in the K-D model,
we suggest that the ground truths of realism are patients' medical images and/or medical knowledge (\ref{item:eval}). The evaluation on realism should not be solely grounded in the latent representation of pre-trained models (or “perception” of the model, as some literature calls). We further discuss this in~\ref{list:fm_assessor_validity} of~\cref{llm}).

\begin{table}[!ht]
    \centering
    \resizebox{\linewidth}{!}{
    \begin{tabular}{p{0.1\linewidth} p{0.15\linewidth}|p{0.15\linewidth}|p{0.3\linewidth}|p{0.3\linewidth}}
    \toprule
    \textbf{Phase} & \textbf{Purpose} &  \textbf{To establish grounding in ...} & \textbf{Evaluation method for the MISyn-enabled downstream task model}& \textbf{Corresponding clinical trial method}\\
    \midrule
\textbf{Phase 0}  & Pre-clinical algorithimc discovery and development& Grounding in the downstream model  & Algorithmic evaluation on efficacy and limitations on hold-out test set & Cell-based (\textit{in vitro}) or animal-based (\textit{in vivo}) experiments\\ \hline
\textbf{Phase 1}  & Algorithmic evaluation on efficacy and limitations 
& Grounding in the K-D model & Algorithmic evaluation on efficacy and limitations on external validation set & Study with tens of healthy human subjects\\ \hline
\textbf{Phase 2} & Primary clinical evaluation on efficacy and limitations & Grounding in the clinical context and K-D model & Evaluation in clinical collaborative settings on efficacy and limitations on external validation set & Study with up to hundreds of patients  \\ \hline
\textbf{Phase 3} &Large-scale clinical evaluation on efficacy and limitations& Grounding in the clinical context and K-D model to approximate the real world & RCT in clinical collaborative settings on efficacy and limitations on retrospective or prospective validation data &Randomized controlled trial (RCT) with hundreds to thousands of patients \\ \hline
\textbf{Phase 4} & Post-deployment monitoring of long-term effects& Grounding in the real world  & Long-term monitoring of the limitations and collaborative efficacy after deployment & Study with any patients receiving the treatment \\
\bottomrule
    \end{tabular}
    }
    \caption{
    The proposed five-phase evaluation paradigm for MISyn-based downstream task models. The 2nd column shows the purpose of each phase. 
    The 3rd column shows that the phased evaluation paradigm gradually grounds new instances from the downstream task model to the downstream model itself, to the K-D model, and to the real world. The grounding is based on study datasets (or study subjects in clinical trials) and study design regarding how the study contexts are controlled. The 4th column shows the evaluation method of MISyn in each phase. Among them, Phases 0 and 1 are algorithmic evaluation that evaluates the efficacy and limitations of the downstream task model. 
Phases 2-4 are clinical evaluation that evaluates the collaborative efficacy and limitations of the model, clinical workers, and clinical outcomes. 
We borrow the word “efficacy” from clinical trial to denote that the evaluation is focused on whether a model achieves the target function it is designed or claimed to do.  
The last column shows the description of the corresponding Phase 0-4 clinical trial.
    }
    \label{tab:phase}
\end{table}
Regarding the evaluation on the performance of downstream task models, to ensure that the MISyn-enabled downstream models are appropriately grounded in the real-world medical phenomena, we suggest using a phased evaluation paradigm by adapting the traditional phased (pre-)clinical study to ML validation~\citep{Park2020,Jin2020,Antoniou2021}. The phased evaluation can set clear methodological standards and expectations in each development stage to gradually ground the downstream models to real clinical settings, and avoid overclaiming and overgeneralization of model performance. Next, we detail the five phases of algorithmic evaluation of MISyn for clinical applications, which is also summarized in~\cref{tab:phase}.

\begin{itemize}
    \item \textbf{Phase 0: Algorithmic discovery.} In technical development, the common model evaluation on the hold-out test set corresponds to the pre-clinical study (Phase 0), usually conducted on animals or cells in laboratory settings. Since this phase is conducted on the hold-out test set that is grounded in the downstream task model, Phase 0 evaluation is to establish grounding of instances from the downstream model in the model itself.
Because of the limitation and discovery nature of this evaluation phase, we cannot claim that the performance on the hold-out test set can generalize to (be grounded in) their real-world performance in clinical settings. 
Doing so is analogous to mistaking the results from pre-clinical study grounded in phenomena on animals or cells as their real-world effects grounded in phenomena on patients.
Instead, the assessment should acknowledge the limitation of its discovery and pre-clinical nature, and warn users that the long-term effect in clinical settings is unknown and should be further investigated in Phase 1 evaluation and above (\ref{item:eval}, \ref{cl:eval}).
\item \textbf{Phase 1: Algorithmic evaluation on efficacy and limitations grounded in the K-D model.} The common model evaluation on external validation datasets corresponds to the Phase 1 study. Starting from Phase 1, the evaluation grounds new instances from the downstream task model beyond the downstream model itself. This is similar to clinical trials in that the concept of a Phase 1 clinical trial is the first phase to test the new drug in human subjects. Since the evaluation is conducted on external validation instances that are grounded in the K-D model, Phase 1 evaluation is to establish grounding of instances from the downstream model in the K-D model.
The study design of Phase 1 evaluation is a highly experimental setting, involving algorithmic evaluation only. 
The Phase 1 evaluation focuses on algorithmic efficacy and limitations to obtain a primary understanding of the scope, strengths, and weaknesses of the downstream task model, which corresponds to the Phase 1 clinical trial to understand the scope, dosage, and side effects of a new drug on humans. 
\item \textbf{Phase 2: Primary clinical evaluation on efficacy and limitations. } 
Starting from Phase 2, the study designs of evaluation become less experimental and closer to the real-world setting. 
To mimic the scenario when the downstream model is embedded in clinical workflow, Phase 2 evaluation assesses the collaborative efficacy and limitations of the downstream task model with clinical users on the external validation dataset. Phase 2 evaluation aims to establish grounding in the K-D model and in clinical collaborative settings. 
\item \textbf{Phase 3: Clinical evaluation on efficacy and limitations to approximate the real world. } Phase 3 evaluation is an advanced version of Phase 2 in that, although this is still in an experimental setting like the previous phases, the evaluation study design is to approximate the key aspects in the real-world clinical settings as much as possible. This usually involves study design that can better mimic the clinical embedding setting of the downstream task model, study methods that can create more solid evidence, such as randomized controlled trial (RCT), and more representative external datasets using retrospective or prospective clinical data. Evidence from the Phase 3 evaluation is also the key to obtaining regulatory approval. 
\item \textbf{Phase 4: Real-world monitoring and maintenance} Like Phase 4 clinical trial, the Phase 4 evaluation is for the ongoing, long-term post-deployment monitoring and maintenance of the MISyn-enabled downstream task model.
\end{itemize}

\item \textbf{Real-world problem grounded motivation. } Last but not least, we should also critically examine our motivations and problem formulation for MISyn and reflect on whether the motivation serves the technical trends and interests only, or is grounded in clinical users' and patients' needs and interests. This is based on the consideration that MISyn can be potentially used to 
mask real-world medical phenomenon such that the masked one is compatible with the capabilities of downstream models,
which creates the illusion of technical progress in the model space that may not translate back to the real world.
Because of this consideration and the previous rationale of the beneficence principle (\ref{p:beneficence} in~\cref{define}), by merely justifying the motivation of MISyn techniques using their prospective technical benefits, such as data augmentation, is far from enough without justifying the association of the motivation with the potential benefits and risks in real-world medical settings (\ref{item:appropriate}, \ref{cl:appropriate}). 
As mentioned previously, the burden of justification is on the technical developers~\citep{10.1145/3636509}. 
\end{enumerate}

\subsubsection{The epistemic benefits of MISyn}

Despite the potential epistemic harms of MISyn on misinformation and its unexpected impacts on downstream tasks, MISyn can also provide epistemic benefits, which is analogous to the epistemic benefits of scientific theories to science: both MISyn and scientific theories are hypothetical spaces, as shown in \cref{fig:worldview}-A; by constructing useful representations of the target phenomenon, they can help us better understand the phenomenon they represent. 
For example, MISyn can be a useful tool to generate new hypotheses by identifying new possible phenomena and expanding existing medical knowledge, although the new hypotheses will need to be verified in real-world medical phenomena to establish their grounding. MISyn samples can also be used to understand the model, such as generating counterfactual explanations to explain black-box AI decisions using MISyn~\citep{cohen2021gifsplanation}.
To utilize the benefits and minimize the harms, we need to critically rethink our existing approaches to the design and evaluation of MISyn techniques, and adapt them towards more ethical MISyn, which we will explore in~\cref{action}.

\subsection{Ethical challenges of foundation model-based MISyn}\label{llm}
In previous subsections, we analyze the limits of MISyn in general. In this subsection, we analyze the ethical challenges of large pre-trained models for MISyn tasks in particular. Large pre-trained model, or foundation model (FM), is defined as ``any model that is trained on broad data (generally using self-supervision at scale) that can be adapted (e.g., fine-tuned) to a wide range of downstream tasks''~\citep{DBLP:journals/corr/abs-2108-07258}. Examples of FM include BERT~\citep{kenton2019bert}, GPT~\citep{chen2020generative}, and CLIP~\citep{radford2021learning}. 
FMs are particularly attractive to MISyn due to their wide range of generative capability from self-supervised pre-training on a broad spectrum of datasets. Meanwhile, FMs pose new ethical challenges to MISyn, as detailed below.

\begin{enumerate}[leftmargin=*,label=\textbf{F\arabic*}]
\item \textbf{Ethical challenges in the epistemic grounding of FM-based MISyn models}:
Grounding its knowledge in the medical datasets and knowledge is a prerequisite for its prospective real-world benefits and ethical benevolence of a MISyn model. However, FMs pose additional challenges to the epistemic grounding of MISyn. Compared to traditional MISyn models\footnote{Here we use traditional MISyn models to denote MISyn models that are not related to or enabled by foundation models. } that mainly ground their knowledge in the target medical knowledge and dataset for the given task, 
FM-based MISyn models have a much larger portion of their knowledge that is grounded in the pre-trained data than the target medical knowledge and dataset.
The FM-based MISyn may not capture the right medical features we aim to generate, and it may be a challenge to fine-tune the model to fully capture the target medical features we want it to grounded in. 
The complicated grounding may also make the assessment of FM-based MISyn models challenging, as the benchmark datasets used in the assessment may not be comprehensive enough to cover the full spectrum of the grounding capability and distribution shift of the FM-based MISyn models~\citep{NEURIPS_DATASETS_AND_BENCHMARKS2021_084b6fbb}. 

\item \label{list:fm_assessor_validity} \textbf{Measurement validity of the FM realism assessor}: 
In addition to the grounding complexity  of FM-based MISyn models, another problem related to grounding complexity is when a FM (or other pre-trained model) is used as an automatic assessor to evaluate realism for traditional or FM-based MISyn models. The FM assessor is used to simulate humans’ perception in assessing how realistic the synthetic images are.
It does so by calculating the distance of test images in the latent representations of the FM assessor using metrics such as Inception Score~\citep{salimans2016improvedtechniquestraininggans}, Fréchet Inception Distance (FID)~\citep{10.5555/3295222.3295408}, or Kernel Inception Distance (KID)~\citep{bińkowski2018demystifying}. However, using FM as a realism assessor is problematic, as detailed below. 

First, if realism was evaluated solely by the FM assessor, the realism evaluation method is invalid. 
The purpose of realism evaluation is to test whether the synthetic images can be grounded in the target K-D model space. As we stated in \cref{sec:eval}, the ground truth knowledge of the realism evaluation are the target medical knowledge and datasets. If FM is used as the sole assessor in the realism evaluation, then the evaluation is grounded in the FM model, and the FM model cannot fully represent the target K-D model according to \cref{axiom1}. This makes the realism evaluation invalid based on measurement theory~\citep{Bandalos2018-sv,10.1145/3442188.3445901}, because the evaluation method fails to achieve its purpose of measuring realism. 
Using FM as the sole realism assessor is an example of “surreptitious substitution” \citep{frank_blind_2024} that surreptitiously substitute the FM model for the target K-D model.  

Second, to use an FM (or other pre-trained model) as a realism assessor, the evaluation method should be conditioned on all of the following: 
\begin{enumerate}[label={\arabic*)}]
    \item The FM should pass all the rigorous tests to establish its grounding in the target medical knowledge and datasets, such that the FM assessor can be a valid approximation for the target K-D model. This, however, only grounds the evaluation method in the FM assessor, not the K-D model, because an approximation cannot replace the original phenomenon according to \cref{axiom1}. Therefore, we need to acknowledge the intrinsic limitation of this evaluation method and accompany this evaluation with other evaluations that can directly ground in the K-D model, as detailed in the next two points. 

\item If the evaluation method includes FM as a realism assessor, the evaluation method should declare the intrinsic limitation of this approach that the FM assessor can only ground the realism evaluation in the FM model space as an approximation, but cannot fully represent the target medical knowledge and datasets that the realism evaluation aims to assess. 

\item In order to truly evaluate realism, the assessment should be accompanied by other evaluation methods that can directly ground the assessment in the target medical knowledge and datasets (\ref{item:eval}). 
\end{enumerate}

\item \textbf{FMs could introduce additional biases}:
Another side effect of FMs is the ethical risk of introducing additional biases in the large-scale pre-training dataset, the pre-training and fine-tuning process, and propagating the biases to downstream tasks~\citep{li2024surveyfairnesslargelanguage}. A recent empirical study shows that ``bias widely exists in FMs for medical imaging'' and ``existing unfairness mitigation strategies are not always effective''~\citep{jin2024fairmedfm}.

\item \textbf{Ethical risks in the data source and training processes}:
Some FMs may use problematic data sources and training processes that pose ethical risks that may not comply with ethical standards in medical settings, such as the undue consent process, copyright infringements, privacy leakage, labor exploitation in data labeling and model training, and the environmental impact of large-scale training processes~\citep{10.1145/3442188.3445922, deng2024deconstructingethicslargelanguage, Ong2024}. 
Given the scale of data and model and given that some FMs may not even be transparent in their data sources and training details, it becomes extremely difficult to scrutinize and audit the data and model, compared to traditional MISyn models.

\end{enumerate}
\part{MISyn and Ethical Practice}

\section{Ethical recommendations of MISyn for technical and non-technical stakeholders}\label{action}

Theoretical analysis should be accompanied by operationalizable methods to effectively implement the ethical principles in MISyn research and development. Building ethical MISyn techniques requires collective efforts within and outside the MIA community. In this section, we provide our inputs of the ethical recommendations to be considered for ethical MISyn practice by summarizing our considerations in Sections~\ref{define} and~\ref{misinfomation}. These recommendations include: 1) a list of ethical practice recommendations to support researchers' and practitioners' design, evaluation, report, and peer review of MISyn techniques (\cref{fig:teaser}); and 2) a list of oversight recommendations for external reviewers, regulators, funding agencies, clinical users, patients, their families, patient groups, and the general public to check if ethical considerations are embedded in the MISyn techniques. Accompanied by other general criteria in AI ethics~\citep{Floridi2018} and research ethics~\citep{10.1093/schbul/sbj005}, the MISyn-specific ethical recommendations can serve as a starting point for the community to open critical discussions on this topic, and to iterate new peer review criteria and practice guidelines with a wide range of stakeholders --- especially with the non-technical stakeholders of the potentially mostly impacted groups consisting of patients, caregivers, and clinical users --- toward more scientifically rigorous and ethical MISyn.
For each item in the two recommendation lists,
we associate it with its rationale by indicating an item number (\ref{item:terminology}--\ref{item:limit} from the ethical practice recommendations, or \ref{cl:appropriate}--\ref{cl:limit} from the oversight recommendations) in parentheses in previous sections.

\subsection{Ethical practice recommendations for technical development of MISyn}\label{action_points}

\begin{enumerate}[label=\textbf{T\arabic*}]

\item \textbf{Terminology}\label{item:terminology}: One of the epistemic limits of synthetic images is that, unlike medical images, synthetic images do not automatically establish grounding in real medical phenomena. Not acknowledging this limit permits the misinformation of MISyn, which is a serious violation of ethics. 
To avoid unethical conduct of misinformation that confuses synthetic images with medical images, the MIA community needs to come up with a clear scope, definition, and unique term for synthetic/generated images in medical settings.
We do not suggest using the term ``synthetic medical image'' because ``medical image'' is a specific term already used in the medical community to denote images captured from patients. Using the words ``synthetic'' (fake) and ``medical'' (real) jointly as the modifier for ``image'' may confuse clinical users and blur the distinction between fake and real. In addition, expressing the meaning of fake images using a three-word phrase will create an additional cognitive load. To equalize the cognitive load when describing fake and real images and make it more intuitive, we suggest using a two-word term ``synthetic image'' (corresponding to its two-words counterpart of ``medical image''), or even one-word term, such as our invented word ``symage'' (corresponding to its single-word counterpart of ``image'' when used in medical context), to describe the synthetic nature of images.

\item \textbf{Technical standard}\label{item:standard}: To avoid the misinformation risk of MISyn, the MIA community needs to set up technical standards to differentiate or label synthetic images from medical images in all the outputs of MISyn techniques, including the report, code, interface, etc., such that technical and non-technical users can be easily informed that the images are synthetic and were not obtained from patients. For example, the synthetic images should be labeled as ``synthetic'' in the image file/folder name, metadata, DICOM field, or using a watermark~\citep{Dathathri2024}. For synthetic images in DICOM format, we suggest adding a new value of ``SYNTHETIC'' to the Content Qualification Attribute (Tag 0018,9004) and label the synthetic image as such. The image metadata should also label other necessary information including the algorithmic source of the synthetic images, and the necessary parameters for generating the synthetic images such as the inputs to the MISyn model, the training data information, and the key evaluation results. And the ``synthetic'' label should always be presented to users when viewing the synthetic images. 

\item \textbf{Usage declaration}\label{item:declare}: If synthetic data are used in downstream tasks, datasets, and models, the usage of synthetic data should be declared, including the component and portion of synthetic and real data, and how the synthetic data were generated and evaluated. This is to avoid the misinformation risk that by default assumes the data are composed of real medical images only.

\item \textbf{Justifiable motivation of MISyn}\label{item:appropriate}: Before proposing or developing MISyn techniques, to ensure that they can potentially serve the ethical principle of beneficence and non-maleficence, researchers and practitioners of MISyn techniques should justify their motivations and problem formulation by answering the following questions:

\begin{enumerate}
\item Is it really necessary to address the motivated problem, and is the framing of the problem valid? Considering the intrinsic limits of MISyn and the significant financial and environmental costs, how does the proposed MISyn technique justify its motivation in comparison to other alternatives, such as no action, non-technical, low-technical, or other technical solutions? 
Are the prospective benefits of the MISyn technique, and their underlying assumptions, justifiable by scientific, ethical, and technical principles, and supported by evidence? 
    \item Whose benefits does the proposed MISyn technique serve? 
    Does the MISyn merely serve for the technical people's pursuit and interests, which could be convoluted with the power of capital~\citep{Steinhoff2022,doi:10.1080/15265161.2024.2377139,10.1145/3531146.3533194}, or does the motivation of MISyn primarily stem from clinical users, stakeholders, and/or public interests and benefits? 
    As we stated in the beneficence principle~\ref{p:beneficence}, ~\cref{epistemic_injustice},  and~\cref{sec:eval}, since the conventionally-formed technical motivations can encode unjust assumptions that discriminate against non-technical people, simply justifying the motivation of MISyn techniques using their prospective technical benefits, such as data augmentation, is not enough without associating the technical motivations with their potential real-world problems supported by evidence.
    \item Who could be potentially harmed, placed at risk, excluded, or overlooked, due to the framing of the problem in this particular way as proposed by the MISyn technique?
\end{enumerate}

\item \textbf{Design assumptions}\label{item:design}: As analyzed in~\cref{setup} and~\ref{sec:risk}, establishing a solid epistemic grounding in medical knowledge and datasets is the prerequisite to legitimize the use of MISyn techniques. In the design of the MISyn techniques, researchers and practitioners should be transparent and specific about their assumptions and epistemic grounding embedded in the MISyn techniques, including what clinical knowledge and assumptions in the training dataset are used in MISyn, and how they are encoded in the model (via the loss function, model architecture, evaluation metrics, etc.). 
Such information can help to assess whether the clinical knowledge and datasets are adequate enough to ensure that the synthetic images approximate the target image distribution, and understand the potential shifts in data distributions and biases between the synthetic images and the target real images. Specifically, researchers and practitioners of MISyn techniques should provide answers to the following questions in the MISyn report:
\begin{enumerate}
    \item What are the key assumptions in the MISyn modeling process? Does the assumption incorporate enough clinical knowledge?
    \item What potential data distribution shifts do the MISyn introduce? 
    \item What potential biases would be introduced in MISyn? Which population could be impacted from the biases? 
\end{enumerate}

\item \textbf{Evaluation}\label{item:eval}: 
The evaluation of MISyn is composed of two aspects: 1) testing the hypotheses of the potential efficacy and benefits of MISyn proposed in the motivation; 2) assessing the risks and harms of MISyn, which is listed in~\ref{item:limit}. While testing the hypotheses of the efficacy of MISyn (i.e., the evaluation on realism) and its benefits (i.e., the performance on downstream tasks), researchers and practitioners should be aware of these problems:

\begin{enumerate}
    \item Is the realism evaluation grounded in the target patients' medical images and/or medical knowledge? As we analyzed in~\ref{list:fm_assessor_validity} of~\cref{llm}), it is invalid to assess  realism solely by a foundation model or other pre-trained model. And such evaluation should be accompanied by other evaluation methods to ground the realism evaluation in the target medical knowledge and datasets, and should acknowledge the intrinsic limitations of using foundation models or other pre-trained models as the realism assessor. 
    \item Do the results on the performance of downstream models overclaim or overgeneralize themselves, such as implying the performance on hold-out test data can generalize to real-world performance, or the algorithmic evaluation on external validation dataset can generalize to collaborative clinical settings? 
    \item\label{item:eval-overclaim} Does the MISyn evaluation report contain unscientific claims or overclaims that could not be backed up by evidence? 
\end{enumerate}

\item \textbf{Limitation analysis}\label{item:limit}: 
Researchers and practitioners should conduct a thorough limitation analysis as the default evaluation routine in MISyn reporting, along with the evaluation of algorithmic efficacy and benefits when proposing or developing a MISyn technique. The role of limitation analysis is similar to that of side effects assessment for medications or medical devices. The goal of limitation analysis is to identify the scope, weaknesses, and failure modes of the MISyn technique, and to acknowledge the limits of the proposed MISyn, so that readers and users can have a holistic understanding of its pros and cons. Specifically, the contents of limitation analysis for MISyn could include the following aspects: 

\begin{enumerate}
	\item \textbf{Assessing the scope, weaknesses, and failure modes of the MISyn}: Similar to the evaluation of the positive aspects of MISyn, researchers and practitioners should conduct equivalently thorough assessments on the negative aspects of the MISyn to understand its scope, weaknesses, and failure modes. For example, a quantitative analysis investigating the potential data distribution shifts and biases introduced in the synthetic data; a qualitative investigation on the failure mode of synthetic data. It should be noted that conducting the limitation assessment and identifying weaknesses of the proposed MISyn should not be regarded as the reason of rejection in peer review (to the contrary, missing it should be considered as a reason of rejection), but should be regarded as a necessary component in the holistic evaluation on the pros and cons of the proposed MISyn. 

\item \textbf{Acknowledging the technique-specific limitations}: Researchers and practitioners should acknowledge the technique-specific limitations in the design assumptions and evaluation methodologies of MISyn, their potential negative impacts, and how future work may mitigate them. For example, acknowledging that specific design considerations can introduce distribution shifts; and acknowledging that the performance evaluation is only grounded in the model and the results could not represent performance in real-world medical settings.

	\item \label{item:intrinsic-limit} \textbf{Acknowledging the intrinsic limits of MISyn}: 
 Different from technique-specific limitations that are specific to the proposed MISyn and may be improved in future work, the intrinsic limits of MISyn are inherent weaknesses compared to real medical images that any MISyn techniques have and are beyond the scope of technological fix. It is important to acknowledge the intrinsic limits of MISyn to avoid the misconception that as technology progresses, the gap between synthetic and real images can become infinitesimal, almost nonexistent, so that we can enjoy the benefits of MISyn with little to no costs or harms.

The intrinsic epistemic limits of MISyn we analyzed include: 1) Unlike medical images, synthetic images do not automatically establish grounding in real medical phenomena, and cannot fully represent the medical images where the MISyn model trained on. 2) The MISyn modeling process inevitably introduces additional distribution shifts and biases to the synthetic data. Despite the possible efforts to mitigate distribution shifts and biases, there always exist distribution shifts and biases that cannot be completely removed. 

In \cref{app:statement}, we provide a template to facilitate the declaration of the intrinsic limits of MISyn.

\end{enumerate}

\end{enumerate}

\subsection{Oversight recommendations for non-technical stakeholders}\label{sec:oversee}

\begin{enumerate}[label=\textbf{O\arabic*}]
    
    \item 
    \textbf{Questioning the appropriateness of MISyn}\label{cl:appropriate}: To counterbalance the dominant power in technology that could be unjust and unethical, it is important to be critical about the utility and appropriateness of synthetic images in medical settings in the first place. Stakeholders and the public can doubt the appropriateness of MISyn by asking the following questions:
    \begin{enumerate}[label=\textbf{O\arabic{enumi}.\arabic*}]
\item What are the end purposes of the synthetic images and the MISyn technique? Does the problem really need to be addressed, or is it a fake problem driven by incentives from dominant power in the technical sector? In comparison to the proposed MISyn solution, are there any other alternatives to address the problem using non-technical, low-technical, or other technical solutions?
\item Whose benefits, interests, and values will the MISyn technique serve, and who (especially the underrepresented social groups) could potentially be harmed, given the fact that the synthetic images cannot provide a full representation of the real medical phenomena? Does the MISyn technique genuinely serve for the benefits of real-world clinical use by providing solid evidence or justification to support it, or it ignores the stakeholders' aspect and is merely for the sake of the technical party's so-called ``technical advances'' or capital increases~\citep{Steinhoff2022} by using synthetic data to improve the model performance, but such performance may not necessarily be convertible to real-world clinical performance?
\item Does the technical party provide enough and balanced information to judge the pros and cons of the appropriateness of MISyn technique and avoid creating a lopsided viewpoint by only emphasizing the benefits? 
    \end{enumerate}
    \item \textbf{Be aware of misinformation of MISyn}\label{cl:misinformation}: It would be unacceptable if the technical party committed misinformation of MISyn, i.e., synthetic images were used 
    without explicitly stating their synthetic nature. This is a violation of ethical principles and destroys trust by intentionally or unintentionally misleading users and stakeholders to mistake the synthetic images as medical images. 
    \item \textbf{Labeling synthetic images}\label{cl:terminology}: Labeling synthetic images as ``synthetic'' is the way to avoid misinformation. Therefore, it is important to check if the technical party has set up mechanisms to ensure that the synthetic images are labeled to be easily distinguishable from medical images. 
    The mechanism for labeling synthetic images as ``synthetic'' should be in place for all the outputs of the MISyn technique, including the model report, the code base, the user interface, etc. 

    \item \textbf{Be vigilant for overclaimed efficacy and benefits of MISyn}\label{cl:eval}: The evaluation on the positive aspects of MISyn usually focuses on two assessments: 1) the efficacy or realism evaluation assesses how realistic the synthetic images are compared to real images. 2) the benefit evaluation assesses the benefits of MISyn to downstream tasks. When interpreting the evaluation results of MISyn, it is important to be vigilant for overclaiming and overgeneralization. Every evaluation has a scope that is confined by the chosen methodology of the assessment, including the hypothesis, task, dataset, evaluation metrics, etc. Overclaiming and overgeneralization happen when the interpretation of results is beyond its evaluation scope. For example, the result on the benefits of MISyn in one downstream task is overgeneralized to other tasks, or the performance on hold-out dataset is overclaimed to be transferable to equal performance on real-world data without supporting evidence from evaluation. Furthermore, we should also be vigilant for unscientific claims in evaluation. For example, it is common to see unsubstantiated claims of performance superiority in ML research. Without proper statistical hypothesis testing, these studies cannot rule out the null hypothesis that observed improvements are simply due to chance.

    \item \textbf{Limitation analysis and disclosure}\label{cl:limit}:
It is important to examine whether the researchers or developers of MISyn conducted limitation analysis and made the required disclosure. The disclosure aims to reveal all information that may influence developing and reporting of MISyn, such as sources of funding and conflicts of interests~\citep{10.1145/3531146.3533194,birhane2022values,stengaard2020}. The limitation analysis is to assess the negative aspects of MISyn, including: 1) a comprehensive limitation assessment on the scope, weaknesses, and failure modes of the MISyn using quantitative and/or qualitative methods; 2) acknowledging the  limitations of the MISyn in its design considerations and evaluation methodologies;
3) acknowledging the intrinsic limits and risks of using synthetic data in comparison to real data. 
    
\end{enumerate}

\section{Instantiation of the recommendations: Case studies}
\label{sec:case_study}

To provide concrete examples of applying the recommendations in MISyn practice, we instantiate the ethical practice recommendations in~\cref{action} in two cases in MISyn research. The two case studies are conducted in the format of a peer review for two recent works published in high impact journals in the MIA field: Case 1 DS3D is a work on traditional MISyn that uses 3D human body model to simulate and generate 2D in-the-wild skin lesion images~\citep{SINHA2024103145}; and Case 2 MedSyn is a work on FM-based MISyn models that uses text-to-image synthesis to generate 3D lung CT~\citep{10566053}. Our example of the full peer review comments according to the ethical practice recommendations are in~\cref{app:case_study}. In~\cref{tab:case_study}, we summarize key findings from the case studies by highlighting the differences between the current practice of MISyn and the proposed more ethical practice according to the recommendations. 

\begingroup 
    \small
\setlength\extrarowheight{2pt}
\setlength\LTcapwidth\textwidth
    \begin{longtable}{>{\raggedright}p{0.15\linewidth}|p{0.40\linewidth}|p{0.40\linewidth}}
        \caption{Summary of the two case studies in~\cref{sec:case_study} showing the differences of MISyn practice between the current and the proposed practice. The current practice is the description of the paper review for the two works of Case 1 DS3D~\citep{SINHA2024103145} and Case 2 MedSyn~\citep{10566053}. The proposed practice is suggestions for ethical improvement for the two cases based on the ethical practice recommendations listed in~\cref{action_points}. }
    \label{tab:case_study}\\
    \toprule
    \endfirsthead
    \endhead
\endlastfoot 
\multicolumn{3}{r@{}}{\em Continued on the next page}\\
\endfoot
    \textbf{Ethical practice recommendations}     & \textbf{Current practice}  & \textbf{Proposed practice} \\
    \hline
\textbf{\ref{item:terminology}, \ref{item:standard}, \ref{item:declare} Terminology, technic-\\al standard, and usage declaration to avoid misinformation}   & 
Both cases clearly label synthetic images as ``synthetic'' in their papers, but do not provide reasonable precautions to prevent misinformation of synthetic images originated from the paper and the publicly-available code base.
& 
In addition to the precautions in the current practice, we suggest 

1) clearly labeling synthetic images as ``synthetic'' in the code base; 

2) alerting the potential misinformation risk of mistaking the synthetic images as medical images in the paper and code base; and 

3) reminding users to label and declare the usage of synthetic data in downstream datasets, tasks, and models. 
\\ \hline
 \multirow{2}{=}{\textbf{\ref{item:appropriate} Justifiable motivation}}    & \textbf{Case 1} One of the main motivations is to generate large-scale dataset. It lacks arguments to associate this technical motivation with particular clinical motivations. 
 Such a motivation can be seen as primarily serving technical advances rather than clinical benefits.

 & \textbf{Case 1} 
 We suggest a clear alignment of 
 the technical motivation with relevant clinical problems, such as wound localization and segmentation when real data is extremely limited.\\ \cline{2-3}
& \textbf{Case 2} 
The presented motivation is focused on 
technical novelty, and lacks justification on how the proposed novel technical features can be associated with and potentially benefit clinical problems. It is not primarily oriented to serve the benefits of potential clinical users. & \textbf{Case 2} We suggest explicitly associating the novel technical features of text-to-image 3D synthesis with potential medical imaging problems, such as 3D surgical simulation with user-customizable inputs, or 3D modeling of airway or vessel stents. \\ \hline
\multirow{2}{=}{\textbf{\ref{item:design} Design assumptions}}    &\textbf{Case 1} It lists specific design assumptions such as modeling parameters, blend loss, and selection of skin diseases. & \textbf{Case 1} In addition to the current practice, we suggest 

1) declaring the important assumption used in modeling, which assumes that skin lesions may not depend on contextual information such as lesion size and body location; this design assumption may not incorporate enough clinical knowledge to sufficiently reflect the contextual information of skin images; 

2) declaring that these design assumptions can introduce additional biases and data distribution shift.\\ \cline{2-3}
& \textbf{Case 2} It reports important design assumptions  on incorporating clinical knowledge to guide image synthesis, disease type and scale, and accuracy of anatomical segmentations.
& \textbf{Case 2} In addition to the current practice, we suggest declaring the important assumption used in modeling and providing evidence for it. The assumption assumes that the medical BERT has learned the target language embeddings to be adequate for encoding CT lung imaging report.
\\ \hline

\multirow{2}{=}{\textbf{\ref{item:eval} Evaluation}}  &   \textbf{Case 1} The main evaluation is conducted on the performance of downstream models.& \textbf{Case 1} We identify the following areas for improvement in the current practice:

1) The realism claim should be backed up with evaluations.

2) The claim on significant performance improvement should be backed up with statistical tests.

3) The reporting should avoid implicit bias that highlights the good performance while downplaying the bad performance. 

\\ \cline{2-3}
& \textbf{Case 2} To assess realism, it conducts both computational evaluation and a user study with radiologists.  & \textbf{Case 2} We identify the following areas for improvement in the current practice:

1) Research ethics board (REB) approval should be reported when using a private dataset or conducting a user study.

2) The use of deception in study design raises ethical concerns. And its potential harms are not addressed or mitigated. \\ \hline

\textbf{\ref{item:limit} Limitation analysis} & Both cases acknowledge the technique-specific limitations. & In addition to the current practice, we suggest 

1) declaring the intrinsic limitations for any MISyn techniques, and 

2) conducting limitation assessment to identify the scope, weaknesses, and failure modes of the proposed MedSyn.
\\
\bottomrule
    \end{longtable}
\endgroup
\FloatBarrier

\section{Conclusion}\label{conclusion}
In this work, to address the ever-increasing needs and requirements for the ethical conduct of MISyn research and development, we provide a theoretical analysis and practical support for ethical MISyn.
Due to space constraints, in this paper, the theoretical analysis and practical support only focus on the essential and common properties of ethical MISyn, such as the identification of several intrinsic limits and the evaluation of MISyn on realism and utility. Future work should critically examine other ethical properties of MISyn mentioned in~\cref{define} (such as fairness, privacy, security, and explainability), conduct ethical analysis and identify task- and application-specific ethical requirements for various MISyn applications mentioned at the beginning of the paper. 
This work is also limited by the particular ethical theories, approaches, background assumptions, and perspectives we choose, which open room for discussion from various other perspectives.
As a primary step towards systematical identification of the epistemic limits and ethical practice in MISyn, future works are needed to iterate on the proposed ethical recommendations in practice to refine them with technical and non-technical stakeholders. 
We hope this work can inspire more works and open discussions in the community to leverage collective efforts in setting up research and development standards for scientifically rigorous and ethical MISyn.

From our analysis, we identify several gaps in the current literature, and future research directions for ethical MISyn could include: 
\begin{enumerate}
    \item Compared to the methodologies on the evaluation of positive aspects of MISyn, the available  methodologies to evaluate the negative aspects of MISyn are not adequate. Future work can develop new assessment standards, methods, metrics, and benchmarks for the limitation analysis and reporting of MISyn that can measure different features of failure modes of MISyn that are clinically relevant~\citep{Burnell2023}, investigate the deviation and biases of the synthetic data from real data, and understand the long-term impacts of using synthetic data in medical settings.
\item Our work analyzed several intrinsic epistemic limits of MISyn. Future works are needed to identify and understand various aspects of the intrinsic limits of MISyn, such as theoretical and empirical investigations of the ``model collapse'' phenomenon~\citep{Shumailov2024,peterson2024aiproblemknowledgecollapse,alemohammad2024selfconsuming,gerstgrasser2024is,shumailov2024curserecursiontraininggenerated} (introduced in~\cref{sec:eval}) in MISyn.
\end{enumerate}

\section*{Author contributions: CRediT}
Weina Jin: Conceptualization, Formal analysis, Writing – original draft

Ashish Sinha: Conceptualization, Writing - Review \& Editing

Kumar Abhishek: Conceptualization, Writing - Review \& Editing

Ghassan Hamarneh: Conceptualization, Writing - Review \& Editing, Supervision, Funding acquisition

\section*{Acknowledgements}
We thank Drs. George Medvedev, Jeremy Kawahara, Ben Cardoen, and Lisa Tang for their helpful comments and valuable discussions.

\section*{Funding sources}
This project is supported by the Natural Sciences and Engineering Research Council of Canada Discovery Grant RGPIN/6752-2020.

\appendix

\renewcommand{\thesection}{H\arabic{section}}
\renewcommand{\thesubsection}{\thesection.\arabic{subsection}}
\renewcommand{\thesubsubsection}{\thesubsection.\arabic{subsubsection}}

\part*{Appendix}

\section{Philosophical and ethical backgrounds}\label{app:bg}
In \cref{sec:assumption} of the paper, we list four background assumptions that set the ethical and philosophical basis for the ethical theories and practice of MISyn in this work. In this section of the Appendix, we introduce the philosophical and ethical backgrounds of these assumptions. 
\begin{enumerate}
    \item \cref{ethics} corresponds to the background assumptions of \ref{b:coproduce} co-production, \ref{b:soutionism} techno-solutionism, and \ref{b:value} the value ladenness of technology. It argues why ethics is indispensable in technical development, and argues one of of the root causes of unethical issues in technology being power asymmetry.

\item \cref{worldview} corresponds to the background assumptions of \ref{b:situated} situated knowledge and \ref{b:nowhere} limited representation. It argues one of the root causes of unethical issues in technology being the problematic worldview of science and technology.

\item \cref{epistemic_injustice} corresponds to the background assumptions of \ref{b: epistemic_injust} epistemic injustice and \ref{b:power} power structure. It argues some of the root causes of unethical issues being power asymmetry and epistemic injustice.
\end{enumerate}

\subsection{Why ethics matters in technical development?}\label{ethics}
The technical development usually focuses on the question of ``how'' to propose and evaluate new technical solutions, while ethics enables us to consider a broad range of issues in techniques, from their upstream motivations (the foundational and usually unquestioned ``why'' question), to their embedded value, to downstream real-world impacts (the ``what if'' question).
A prevailing belief is that as long as the technical people do well in their own jobs, they are doing social good by pursuing technical advances~\citep{birhane2022values} and can be exempted from external ethics regulations of their activities. 
However, power --- the technical professionals' discretion --- without check can easily be abused.
Historically, it is exactly the same thought of healthcare professionals that led to bioethical scandals and public distrust, such as the infamous U.S. Public Health Service Untreated Syphilis Study at Tuskegee~\citep{jones1981bad}, among many others~\citep{Beecher1966}. 
In systematic disregard for ethical issues, elite medical researchers purposely injected syphilis, hepatitis viruses, or cancer cells, without consent and without treatment, into the poor, elderly, black people, incarcerated people, or mentally defective children, in the name of pursuing medical knowledge and progress for ``the greatest good for the greatest number'' using ``a utilitarian calculus''~\citep{rothman_strangers_1992}. These were regarded as common practice in the medical community for three decades after World War II ``without either scrutiny or sanction''~\citep{rothman_strangers_1992}: the research were large projects funded by federal offices, the U.S. military, National Institutes of Health, or drug companies; the results were published in prestigious medical journals; and the principal investigators ``[became] national leaders in medicine, chairmen of major departments, and winners of major awards, often for the very research'' that seriously violated ethics~\citep{rothman_strangers_1992}.
Considering the same history is happening nowadays, where marginalized social groups are suffering from systematic algorithmic discrimination, injustice, and exploitation~\citep{neda_surrogate_2019,gray_ghost_2019,williams_exploited_2022}, while these very techniques are celebrated in the AI community as major technical achievements~\citep{Gebru2024},
there is no guarantee that technological professionals can do better than healthcare professionals, whose practice is already required to uphold specific ethical standards for thousands of years.

The above historical example shows excluding ethics from research and practice can cause harmful consequences and endanger the integrity of a field. Theoretically, such a notion that \textit{technical development can be exempted from ethics regulations as long as we technical people do our jobs well} is pervasive because it is based on two assumptions in the technical community: 
\textbf{Assumption (i) Technology is isolable}: 
(ia) Technical development can be abstracted and isolated from, or cannot control, its upstream motivations and downstream social impacts~\citep{matthewman_technology_2011}; or (ib) Even if there are ethical issues occurred during technical development, they can be fixed technically within the technical community~\citep{noauthor_deep_nodate}. \textbf{Assumption (ii) Technology is value-neutral}: Technical development itself is value-neutral~\citep{matthewman_technology_2011}, or its only embedded value is for technical advances that can eventually serve for social good, thus technical development is free from ethical and value judgments. However, the two assumptions are myths that are not supported by evidence, which we will detail below.

For assumption (ia), just because technical people try to isolate our activities from the social world and evade our responsibilities does not necessarily indicate that technical activities are in fact isolable and responsibilities are evadable~\citep{matthewman_technology_2011}. Given our current reality that technology and society are deeply entangled with each other, it is impossible to isolate one from the other even if we want to. Therefore, this assumption needs to be updated to fit the complex phenomena in our contemporary lifeworld that technology ``shapes and is shaped by the social world''~\citep{BoenigLiptsin2022}, and ``society and technology as co-productions''~\citep{matthewman_technology_2011}.
Assumption (ib) advocates for a techno-solutionism mindset that any problem can be solved technically~\citep{TechnologicalFix2024}. Just because we have a hammer does not justify that we can treat every problem as if it were a nail. This is a cognitive bias that overestimates the capability of a familiar tool~\citep{noauthor_law_2024}. Given a problem caused by technical development, shouldn't the logical way to solve it is to first diagnose the root causes of the problem, and then identify treatment that can target the root causes? There is no guarantee that the treatment solution will always fall into the realm of technical toolbox. For ethical issues occurred in technology, ethics knowledge is always needed for precise diagnosis, treatment, and prevention. 

For assumption (ii), if one claims that the technical development is to serve for social good, then who have the right to define what is social good? Whose values and interests are represented, and whose are ignored in the definition? The right to define technical objectives should be in the collective power of civil society, not in the hands of few technical elites. Scholars have pointed out that this central assumption in AI development is related to the eugenic ideology~\citep{Gebru2024,Cave2020,wigginsHowDataHappened2023,chunDiscriminatingDataCorrelation2021}, where ``a small and largely homogenous group of people have the legitimate ability to decide what is `good' for `humanity'.''~\citep{Burrell2024} 
Furthermore, claiming technology to be value-neutral or value-free is meaningless, because being a tool, the development and application of technology cannot be isolated from its objective and usage context, which reflect values on what are important to pursue~\citep{verbeekMoralizingTechnologyUnderstanding2011a}.
The ability to define and influence others on what are important defines power~\citep{Fricker2007_testimonial}. 
Here, power is a Foucaultian concept that ``operates through impersonal institutions that no one particular person may have control over''~\citep{Burrell2024}.
The dominant power structures ``lay out paths of least resistance that shape how people participate''~\citep{Johnson2014-gj}. For example, the paths of least resistance set the research and development agenda and incentives of what values, ideologies, and worldviews to embed in technologies~\citep{birhane2022values, Gebru2024, Burrell2024, neda_surrogate_2019}; the dominant power structures also shape the technical criteria and design choices accordingly that mainly reflect the technical agenda rather than real-world needs, such as clinical needs~\citep{Reinke2024,JIN2023102684}. The latent power factor implicitly influences how technical people choose and frame their problems and assess their techniques on whose interests are reflected in technology and whose interests can be ignored to gain career advancement in the technical field~\citep{10.1145/3531146.3533194, Burrell2024Automated, Ahmed2024,Newfield2023}. This power mechanism alerts us that merely relying on people specialized in AI ethics technologies (fairness, transparency, interpretability, etc.) to recognize, ``fix,'' and defend against the pervasive penetration of unjust power structures towards ethical technology is insufficient. It should be the responsibility of every people in the technical community to conduct research and development ethically as part of our professional code of conduct.

\subsection{Where do unethical problems of technologies come from? Different worldviews of science}\label{worldview}

In the last subsection, we argue the importance of conducting ethical research and development of technologies, and diagnose that the partial root cause of unethical technologies lies in the unchecked and imbalanced power dynamics in the technical community. The power factor intertwines with different philosophies and worldviews of how we understand science and technology. The dominant philosophy and worldview legitimize and defend for the dominant unjust power structures, which eventually lead to unethical technologies. Therefore, to establish a precise diagnosis of the unethical issue in technology, it is necessary to trace the problem in the philosophy and worldview of science and technology.

\begin{figure}[!ht]
    \centering
    \includegraphics[width=\linewidth]{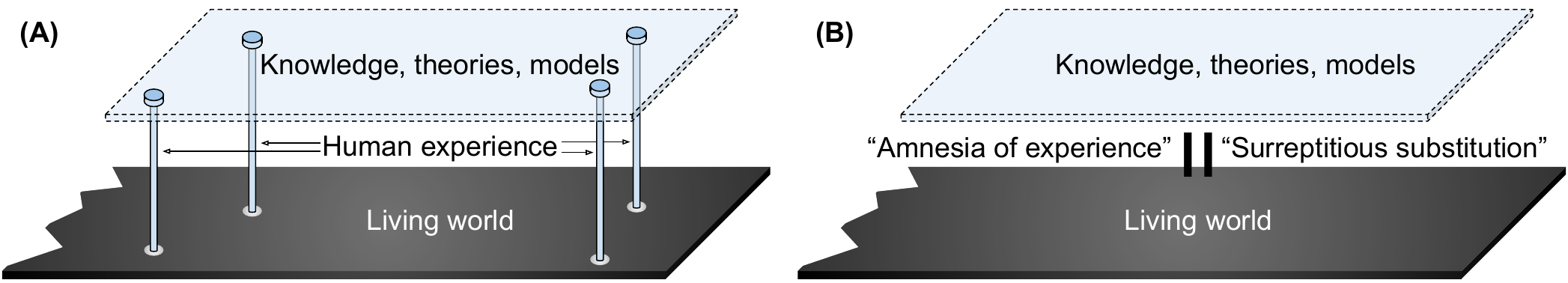}
    \caption{
    The two contrasting worldviews of science and technology. \textbf{(A) The normal worldview}: it visualizes that science and technology (including computational models) are built upon human experience. Human experience also provides the ``epistemic footing''~\citep{frank_blind_2024} for their validity and applications in the real world. 
   \textbf{ (B) The blind spot worldview}: it visualizes how the epistemic footing (human experience) become forgotten (``amnesia of experience''~\citep{frank_blind_2024}) when we have built floors (science and technology) on top of it. It wrongly assumes that science and technology equal the reality and can fully represent the reality (``surreptitious substitution,'' or ``confuses the map with the territory''~\citep{frank_blind_2024}), ignoring the complex and irreducible nature of the real world. Doing so also ignores that we should establish new epistemic grounding for verification before we can confidently apply the computational models in the real world. 
    }
    \label{fig:worldview}
\end{figure}

Frank, Gleiser and Thompson diagnosed the key pathologies of the philosophy and worldview in science that most people take for granted (Table~\ref{tab:blindspot}), and name them the blind spot\footnote{The content of this subsection is mainly a summary of their book \textit{The Blind Spot: Why Science Cannot Ignore Human Experience}~\citep{frank_blind_2024}. We refer interested readers to this book to unpack detailed arguments on the blind spot worldview.}, which is ``the failure to see direct experience as the irreducible wellspring of knowledge''~\citep{frank_blind_2024}. They also diagnosed the manifestation of these pathologies in AI, which they named the computational blind spot (Table~\ref{tab:blindspot}). The pathologies also manifest in MISyn, as we discuss in the paper~\cref{harm}. To illustrate the idea, we contrast two worldviews of science and technology without and with the blind spot in~\cref{fig:worldview}. 

\cref{fig:worldview}-A shows how science and computational models work: scientists or engineers first abstract knowledge, theories, or computational models from concrete human experience in the living world (visualized as the ground in the figure). The human experience, visualized as pillars in the figure, ``is a precondition of observation, investigation, exploration, measurement, and justification''~\citep{frank_blind_2024} that forms data instances and scientific experiments.
The knowledge, theory, or model, residing in the abstract space (the space above the ground in the figure), forms a finite space (the first floor in the figure) confined by the controlled conditions in scientific experiments or assumptions in models. A theory or model can extend itself within the abstract space by conducting symbolic inference (logically or mathematically) to generate new hypothesis (we later name it the symbolic authenticity). However, the newly inferred regions in the abstract space, if not supported by evidence of prior human experiences, could not extend their authenticity automatically in the living world. That is why we need to conduct new scientific experiments to test the new hypothesis, or test the predictive models in the real world to acquire their performance, such that to establish grounding of the new hypothesis or models in reality (we later name it the grounding authenticity). By building abstract theories and models and grounding them back to reality, we enrich our understanding of our living world using science and technology. In this iterative process, both the grounding with human experience and the abstract scientific knowledge or models are valid and necessary. Human experience provides the ``epistemic footing'' that precondition the building of abstract scientific theories and models~\citep{frank_blind_2024}. There is no hierarchy of one over the other.

\begin{table}[!ht]
    \centering
    \resizebox{\linewidth}{!}{
    \begin{tabular}{>{\raggedright}p{0.2\linewidth}|p{0.4\linewidth}|p{0.4\linewidth}}
    \toprule
    \textbf{Pathologies of the blind spot} & \textbf{Definition} & \textbf{Computational blind spot}\\
    \midrule
\textbf{1. ``Surreptitio-\\us substitution''} & ``The replacement of concrete, tangible, and observable being with abstract and idealized mathematical constructs,'' i.e., ``confuses the map with the territory.'' &  ''Substituting computational states for the imprecise and fluid everyday world'' and ``substituting computation for genuine embodied intelligence.''\\ \hline
\textbf{2. ``The fallacy of misplaced concreteness''} & ``The error of mistaking the abstract for the concrete. It underlies the surreptitious substitution.'' & ``Treating abstract computational models as if they were concretely real.'' \\ \hline
\textbf{3. ``Bifurcation of nature''} & ``They divide nature into what exists externally and objectively, and what is mere subjective appearance or exists just in the perceiver’s mind.''
 & ``Human intuition (or the experience of meaning or relevance realization) is a subjective epiphenomenon of objective brain computations.'' \\ \hline
\textbf{4. ``The amnes-\\ia of experien-\\ce''} & ``This happens when we become so caught up in surreptitious substitution, the fallacy of misplaced concreteness, and the reification of structural invariants that experience finally drops out of sight completely.'' &  ``Forgetting that human experience and biases lie deeply sedimented in AI models'' and ``belies AI’s fundamental dependence on physical nature and socially organized, collective human knowledge.''\\
\bottomrule
    \end{tabular}
    }
    \caption{
    The blind spot worldview on science and technology, and its manifestation in AI as the computational blind spot. Only pathologies related to the computational blind spot are listed here, and the full list of pathologies is in~\citep{frank_blind_2024}. The quotes are cited from~\citep{frank_blind_2024}. Refusal of these pathologies provides support for \cref{axiom1} in~\cref{setup}.
    }
    \label{tab:blindspot}
\end{table}
\FloatBarrier
In contrast, \cref{fig:worldview}-B illustrates the blind spot of the mainstream but problematic worldview of science and technology. After acquiring the abstract knowledge, theories, or models, we ``get so caught up in the ascending spiral of abstraction and idealization'' and ``become so captivated by the spectacular success of science that we've forgotten that direct experience is science's essential source and constant support''~\citep{frank_blind_2024}. The blind spot worldview regards the abstraction as more real or authentic than the concrete. Doing so leads to an unnecessary and unscientific hierarchy that the abstracted knowledge can dominate over the concrete human experience and nature. This hierarchical worldview legitimizes science and technology activities that relegate human experience and nature to pure resources to control over, and eventually cause ethical and societal crisis, such as algorithmic discrimination, human rights violation, labor exploitation, and environmental crisis. 
The hierarchical worldview is further justified by ``a loaded and unnecessary metaphysical assumption about what the world is like outside the range of our ability to construct and test predictive models. The assumption is that how things behave in tightly controlled and manufactured environments should be our guide to how things behave in uncontrolled and unfabricated settings. [...] It’s based on generalizing from a small number of cases where we do have successful predictive models to a vastly larger number of cases where we do not, and arguably cannot, have this kind of knowledge, because the world outside the [scientific] workshop is too entangled and complex''~\citep{frank_blind_2024}.
In addition to the above unnecessary hierarchy and unrealistic generalization assumption, for computational blind spot, there are two aspects that prevent us from seeing the limits of computational models. ``One is surreptitiously substituting computational models for the everyday world; the other is failing to see how we are led to remake the world so that it gears into the limitations of our computational systems''~\citep{frank_blind_2024}. Understanding the philosophical flaws in the computational blind spot worldview permits the analysis of the limits of MISyn in the paper~\cref{misinfomation}.

\subsection{Epistemic injustice and the silenced harms}\label{epistemic_injustice}

In the last subsection, we point out another root cause of unethical issues that lies in the problematic philosophy and worldview. The blind spot worldview illogically overvalues scientific theories and computational models and undervalues human experience in knowledge. This fulfills the definition of epistemic injustice in ethics, which ``wrongs someone in their capacity as a subject of knowledge, and thus in a capacity essential to human value''~\citep{Fricker2007}. 
We will discuss two cases of epistemic injustice related to MISyn. Case 1 is placing humans in a subordinate position with respect to computational models in knowing and reasoning, which is related to the unethical conduct of using MISyn-enabled technologies to replace, deskill, or degrade healthcare workers; and Case 2 is placing some human groups in a subordinate position with respect to other human groups in their contribution to knowledge, which is related to the intrinsic epistemic limits of MISyn that MISyn could never fully replace the medical phenomena and knowledge contributed by some human groups.
Both cases cause ``discriminatory diminution of an agent's epistemic status'' and silenced harms to the subordinate groups ``as knowers, interpreters, and sources of testimony''~\citep{Symons2022}. 

For Case 1, we argue that there is an epistemic discrimination against humans in technical development. A prominent example at the macro level is that the mainstream AI objective is fixated on replacing humans~\citep{grace2024thousandsaiauthorsfuture}. While it is totally legitimate for individual researchers and practitioners to freely pursue any valid AI objectives, it is unhealthy for the whole field to fixate on a narrow, rather than diverse, set of objectives. Given that other alternative AI objectives do exist that emphasize a collaborative, rather than competitive, relationship between AI and humans, and may potentially lead to better outcomes for both technical development and human flourishing~\citep{doi:10.1177/26339137231222481,jin2024plausibilitysurprisinglyproblematicxai}, framing the AI objective as the inevitable replacement of humans without critical justifications~\citep{Gebru2024, Goodlad2023} is unreasonable, and denies humans' epistemic capacities (\ref{item:appropriate}, \ref{cl:appropriate}). It further reflects the ideology of computational blind spot and unjust power structures in the AI field~\citep{doi:10.1080/15265161.2024.2377139}.
Another example at the micro level is that many AI techniques are proposed with the motivation to overcome human errors~\citep{Lysen_2023}. While it is worthwhile to mitigate human errors, this motivation unreasonably assumes that there exists ideal AI models that can transcend humans' epistemic imperfections and exhibit epistemic superiority to humans in all aspects\footnote{Again, this reflects the blind spot worldview, as Frank, Gleiser and Thompson addressed, ``Scientific knowledge isn't a window onto a disembodied, God's-eye perspective. It doesn't grant us access to a perfectly knowable, timeless objective reality. [...] Instead, all science is always our science, profoundly and irreducibly human, an expression of how we experience and interact with the world''~\citep{frank_blind_2024}.}~\citep{GONDOCS2024102769}.
It thereby ignores the fact that AI infrastructures and algorithms are built upon human knowledge, judgment, and experience, and could inherit human imperfections and introduce new errors, such as biases in the dataset and algorithms (\ref{item:design}). This epistemic injustice is manifested in how technologies are assessed: the community mainly focuses on evaluating the benefits of algorithms such as performance superiority; but lacks methods, technical standards, and awareness to analyze the limitations, applicable scopes, environmental impacts, and potential issues of the proposed technique as the evaluation defaults~\citep{Burnell2023,HernandezBoussard2020}. 
This contrasts with research practices in other scientific fields and engineering standards, where both strengths and limitations, warnings, and scopes are reported (\ref{item:limit}, \ref{cl:limit}). By merely highlighting human limitations and algorithmic strengths while neglecting human strengths and algorithmic limitations, this approach underlines the epistemic discrimination against humans in the AI field (\ref{item:eval}, \ref{cl:eval}).

For Case 2, the fallacy of the epistemic injustice is that it wrongly assumes that people have unequal epistemic capacities or contributions. For example, it wrongly assumes technical people have more epistemic capabilities than non-technical users in designing a MISyn model. 
This reflects a common prejudice in software development that ignores the intended users' requirements~\citep{DBLP:journals/corr/abs-1712-00547}, and could lead to the failure to incorporate non-technical users' clinical knowledge and experience into the MISyn model (\ref{item:design}). Another example is, for diseases or phenomena that are not modeled or captured in the MISyn models (due to various reasons such as the patients' socioeconomic status or their access to healthcare resources), these phenomena are wrongly assumed to not exist. This prejudice does not recognize the lived experience and sufferings from patients, and ignores the epistemic contribution to the accumulation of medical knowledge from patients with such diseases or phenomena~\citep{Carel2014} (\ref{item:limit}, \ref{cl:limit}). 

The ethical consequences of epistemic injustice are twofold: 1) For the oppressed people, their epistemic capacities as knowledge providers and reasoners, and thus their human value, are undermined. The oppressed are assumed a prior inferiority regarding their epistemic status, making it difficult for them and others to understand the harm being done to them, let alone to be equipped with epistemic equality and resources to justify for themselves.
2) For the community as a whole, it loses the opportunity to access knowledge and truth from the oppressed~\citep{Fricker2007}. These harms create a silencing effect to both the oppressed and the broader community. Considering that these harms usually happen in the institutional context, where the technologies are embedded and encoded with epistemic injustice in different ways detailed in the above examples, it creates an impersonal power in which injustice happens not because of some bad actors or malicious intent, but as ``a consequence of often unconscious assumptions and reactions of well-meaning people in ordinary interactions, media and cultural stereotypes, and structural features of bureaucratic hierarchies and market mechanisms --- in short, the normal processes of everyday life''~\citep{youngJusticePoliticsDifference1990}. It further illustrates how technology acts as a mediator~\citep{matthewman_technology_2011}, shaped by power structures, which in turn shapes and perpetuates unjust social relationships that lead to unethical consequences~\citep{BoenigLiptsin2022}.

\section{Details of the case studies}\label{app:case_study}
In this section of the Appendix, we present the details of the two case studies in ~\cref{sec:case_study} of the paper. The cases demonstrate how to apply the ethical practice recommendations and conduct critical ethical research, development, reporting, and review practice towards ethical MISyn. 
The case studies are conducted on two recent works published in high impact journals in the MIA field on traditional MISyn and FM-based MISyn models, respectively. 
For each paper, we first summarize the paper and then conduct our ethical analysis based on the ethical practice recommendations  in~\cref{action_points}. 
The quotes appearing in this section, if unspecified of their reference, are directly quoted from each paper.
We use works, including our own, not to impute to individual researchers and practitioners, as these issues are quite common in the literature and are not limited to the analyzed papers.
Instead, we aim to urge systematic changes in the community to adapt to more scientifically rigorous and ethical practice in research, reporting, and peer review.

\subsection{Case study 1: Traditional MISyn}
We use our recent work on traditional MISyn as the case study: ``DermSynth3D: Synthesis of in-the-wild annotated dermatology images''~\citep{SINHA2024103145}. 

\textbf{Paper summary}

This is a simulation-based MISyn work. The motivation is to tackle the data scarcity issue in skin image analysis tasks ``including a small number of image samples, limited disease conditions, insufficient annotations, and non-standardized image acquisitions.'' 
The proposed DermSynth3D (DS3D) framework generates 2D in-the-wild skin lesion images with dense annotations for skin, lesions, bounding boxes around lesions, body-part segmentation, depth maps, and other 3D scene parameters from 3D human body models. DS3D crops the 2D skin lesion from skin lesion datasets, blends them onto the texture maps of the 3D human scans, and generates multiple 2D views with dense annotations from the 3D bodies. 
Using 50 manually annotated 2D skin images from the Fitzpatrick17k dataset, the evaluation assesses the performance of downstream models trained on data synthesized by DS3D by studying the effects of mixing different portions of synthetic and real images on two tasks of bounding box detection and semantic segmentation. 

\textbf{\ref{item:terminology}-\ref{item:declare} Terminology, technical standard, and usage declaration to avoid misinformation}

To avoid misinformation of MISyn, the synthetic images and their downstream models should be clearly labeled to distinguish them from medical images. Although DS3D did not confuse the synthetic images with real images by clearly labeling the synthetic images as synthetic (such as in their Figures 3 and 6), and labeling the downstream models that utilize synthetic images (such as in Figures 10 and 11), it did not set up adequate mechanisms to prevent misinformation originating from its paper and code base. For example, when generating synthetic datasets, the publicly-available code base did not explicitly label the images as synthetic in either the image file/folder name or the metadata\footnote{In the code base, although the folder name for the generated synthetic images is customizable and the default folder name is ``blended\_lesions,'' this naming convention did not explicitly state the synthetic nature of the generated images outside the context of the code base when the generated images are used as a standalone dataset. The comment for customizing the folder name did not alert users that the image file/folder name should be explicitly labeled as ``synthetic.''}. The paper and the code base did not alert the potential risks of mixing synthetic with real images, including the misinformation risk of polluting medical image datasets and the learning of downstream models. The paper and the code base also failed to remind readers and users that the usage of synthetic images should be clearly labeled and declared. 

\textbf{\ref{item:appropriate} Justifiable motivation of MISyn}

In the contribution section of the DS3D paper, one of the main motivations of DS3D that uses 3D human models to generate unlimited annotated skin images is legitimized by ``a lack of a large-scale skin-image dataset that can be applied to a variety of skin analysis tasks.'' 
Since the authors did not provide their chain of reasoning to draw logical inferences from the benefits of large-scale datasets to their potential utility in medical imaging applications, and from the problem of lacking large-scale real image datasets to the proposed synthetic image solution, readers have to make two reasonable assumptions to close the gaps in the above chain of reasoning for this motivation:
1) it assumes that large-scale real data is a merit in itself (but the authors failed to provide evidence for it); 
2) it surreptitiously substitutes the benefits from large-scale synthetic images for the benefits from large-scale real images (cf. \cref{axiom1}). 
In fact, the evaluation results from the DS3D paper provided empirical evidence that large-scale synthetic data may not always be beneficial, as ``the performance of the model converges after the addition of 400 synthetic training images and increasing them beyond 1000 did not significantly increase the performance.''

Given that the authors did not explain how the technical motivation (generating large-scale synthetic data) can be associated with clinical motivation (being potentially beneficial to clinical users and patients), it is reasonable to deduce that either the authors failed to perform their responsibility to clearly declare all their motivations for a paper, or the authors did not have clinical application success as their motivation. In either case, the reader may speculate about the authors' objective: 
is the technical motivation for large-scale dataset only for the technical’s sake\footnote{Regardless of the potential harms of not aligning the technical pursuit with clinical problems as we have pointed out in T4.2, this technical objective that large dataset is always better is unreflective in itself, see~\citet{Li2023}.}? Failing to link the proposed technique to a clinical motivation may serve the interests and benefits of the technical researchers and practitioners to advance techniques, publish papers, or gain career advancement, but such unjustified and convention-following motivations (`large datasets are better') may harm clinical users and patients. Instead of justifying MISyn based on vague and unreflective motivations, 
a better way is to correspond the motivation and problem formulation to the hypotheses that will be tested in the evaluation, 
such as motivating the utility of DS3D in wound localization and segmentation when real data is extremely limited.

\textbf{\ref{item:design} Design assumptions} 

By pasting the cropped skin lesion regions to the body parts of the 3D human models, DS3D held a strong assumption about the diseases it models: the skin lesions are context-invariant that may not depend on contextual information, such as body location, size of the lesion, skin tone, the number of lesions and how they cluster, patient's information (age, gender, symptoms, etc.), the camera perspective and lighting condition on the original real image, etc. In fact, skin lesions are usually context-dependent, as the authors stated elsewhere in the evaluation section on ``Predicting body parts from 2D images'':  ``Understanding where the skin condition is on the body may be an important factor in determining likely skin conditions.'' The authors did not declare in their report this important context-invariant assumption used in modeling, and did not mitigate the gap between the context-dependent nature of skin images and their modeling assumption. Mitigating the gap will require incorporating necessary clinical knowledge into the modeling process for the given task, such as modeling the location distribution of different skin diseases. Missing these points indicates the work is inadequate in declaring design assumptions and incorporating clinical knowledge.

Although the authors declared their specific design choices of modeling parameters, blending loss, and selection of skin diseases as their limitations: ``we only blended skin conditions that could be confidently manually segmented, and hence, did not include diffused skin disease patterns such as acne,'' they failed to explicitly declare that these design choices can introduce additional biases and data distribution shift in the MISyn modeling process.

\textbf{\ref{item:eval} Evaluation} 

Although DS3D claims to generate ``photo-realistic 2D dermatological images,'' it did not provide any evaluation, neither quantitatively nor qualitatively, on this realism claim. Despite the simulation nature of the DS3D technique and that there is no ground truth of real skin images to verify the realism, to back up this important claim and to ground the realism evaluation in the target medical knowledge and datasets,
 the authors could at least conduct qualitative study with dermatologists to assess the alignment of the synthetic images with clinical knowledge.  

The main evaluation of DS3D is on the performance of downstream models. However, the evaluation contains overclaims: first, the authors claimed in the abstract that DS3D was evaluated ``on various dermatology tasks,'' whereas only two tasks of bounding box detection and semantic segmentation were evaluated. Both tasks belong to the same task family of segmentation, and the bounding box of lesions can be derived from semantic segmentation. Other synthetic functions of DS3D intended for other dermatological tasks, such as the synthesis of depth maps and 3D scene parameters, were not evaluated. Second, the claimed significant improvements of performance were not backed up by statistical tests to reduce the likelihood that the improvements were purely due to chance, such as the claim ``[w]e can see in Fig.10 that augmenting the entire real training dataset with synthetic images significantly improves the wound detection performance.'' Third, the authors provided imbalanced claims that were biased towards highlighting the good performance while downplaying the bad performance. After noting that the additional synthetic images did not improve performance, the authors provided the unjustified explanation that ``this maybe partly application-dependent.'' However, this explanation can be equally used to excuse bad performance as well as to explain away good performance. 

\textbf{\ref{item:limit} Limitation analysis}

The limitation analysis is to document the limitations and conduct a thorough analysis to assess the potential side effects. In the limitation part of the DS3D paper, the authors failed to acknowledge that 1) the synthetic images are not grounded in real medical phenomena and should not be mistaken as real images; 2) there are additional biases introduced by the MISyn model due to design choices on modeling parameters, blending loss, and lesion selection; 3) the proof-of-concept nature of the work (pre-clinical study) makes it not suitable to be directly applied without verification (evidence from phase 2 and 3 clinical studies) in settings involving humans. 

For the quantification of the side effects on distribution shift and biases, although the work acknowledged the data distribution shifts between the synthetic and real image domains and between different semantic contents, the evaluation did not thoroughly quantify the data distribution shifts and the impacts of biases on downstream models' performances using representative benchmarks and instance-by-instance reporting~\citep{Burnell2023}.

\subsection{Case study 2: FM-based MISyn}
Since our research group does not have published works on FM-based MISyn at the time of drafting this paper, we need to select another work for the case study. The selection process is described as follows: First, to ensure that the two case studies have a good representation from different venues and tasks, we excluded the journal in the first case, and targeted equivalent venues of high-impact journals in the medical image analysis field, and the top-tier conferences in the field of AI or computer vision (CV). Within the selected venues, we then searched for papers in the last two years using keywords of synthesis/synthetic/synthesizing/generation/generative/generating and transformer/diffusion/diffusive/foundation/large, and added the additional keywords of health/medical when searching the general AI/CV venues. We also referred to papers mentioned in recent surveys on FM in medicine~\citep{10847310,AZAD2024103000,ZHANG2024102996,KAZEROUNI2023102846}. From the screened works, we scanned the full-text papers and excluded works that do not involve pre-trained FM, and excluded works on skin image analysis to avoid duplicate task with the first case. 
The case selection process was conducted in August 2024. Among papers that fulfilled the inclusion and exclusion criteria, we selected one recent work: ``MedSyn: Text-guided Anatomy-aware Synthesis of High-Fidelity 3D CT Images''~\citep{10566053}, because this work contains a comprehensive evaluation and reporting, and the FM was pre-trained on medical dataset compared to other works in which the FMs were only pre-trained on non-medical dataset.

\textbf{Paper summary}

This work aims to generate high-resolution 3D lung Computed Tomography (CT) images guided by text prompts and anatomical segmentation masks. It is motivated by the recent technical advances in generative AI such as diffusion models and the technical novelty of ``text-guided volumetric image generation techniques for medical imaging.'' 
The proposed FM-based MISyn model, MedSyn, is a pipeline model consisting of three sub-models: it first uses a pre-trained and fine-tuned language model of medical BERT to encode radiology reports; the text embedding guides a low-resolution 3D diffusion model to generate low-resolution CT images as well as anatomical segmentation masks of airways, vessels, and lung lobes; these outputs are fed to a super-resolution 3D diffusion model to generate high-resolution 3D CT images. 
The evaluations on realism are conducted using computational assessments and a user study with 10 radiologists in comparison with baseline generative models.

\textbf{\ref{item:terminology}-\ref{item:declare} Terminology, technical standard, and usage declaration to avoid misinformation}

Similar to the analysis in Case 1, this work did not provide any reasonable precautions to prevent misinformation of synthetic images originated from its report and publicly-available code base, including: failing to label the generated images as ``synthetic'' or ``generated'' in its code base; failing to alert the potential misinformation risk of mistaking the synthetic images as medical images in its report and code base; failing to remind users to label and declare the usage of synthetic data when using the proposed MISyn work in downstream datasets, tasks, and models.

\textbf{\ref{item:appropriate} Justifiable motivation of MISyn}

Similar to the analysis in Case 1, MedSyn only mentions technical novelty, and lacks justification on how the proposed novel technical features of MedSyn can be associated with and potentially benefit the clinical problems. The motivation in the introduction and related works is based on new technical advantages of text-to-image generative AI and the limitations of existing technical solutions. Although the authors mentioned in the abstract that ``the advancements in image generation can be applied to enhance numerous downstream tasks,'' and listed several application examples in the discussion as future work, such as for data augmentation, model explanation, and generating diverse samples for model audit, these are general applications that other MISyn techniques or real data can be applied to as well. It is unclear how the proposed novel technical features of MedSyn (text and anatomical masks guided, 3D image generation) can be motivated by and potentially enable unique medical imaging applications.

For MedSyn, it is not difficult to identify potential medical imaging problems for which the novel features may be suitable\footnote{The technical community also calls for a paradigm shift for application-driven innovations inspired by real-world challenges, rather than methods-driven innovations only. This could increase social and technical impacts, advance ML research as a whole, and diversify research directions~\citep{pmlr-v235-rolnick24a,Rudin2013,10.5555/3042573.3042809}.}, such as 3D surgical simulation with customizable pathologies guided by users' text input or mask annotations, or facilitating 3D modeling of airway or vessel stents. Ignoring the motivation from medical application perspective indicates that the work is not primarily oriented to serve the benefits of potential clinical users, but for the authors' own interests to follow the mainstream values in the technical community. The ill-motivated techniques that are not tailored to clinical needs could potentially put clinical users and patients at risk. 

\textbf{\ref{item:design} Design assumptions} 

By using the pre-trained and fine-tuned language model of medical BERT to guide the generation of low-resolution 3D CT lung images, the authors utilized an important design assumption that medical BERT encodes necessary language embeddings to be capable to perform the target task. But the authors did not report evidence to support this assumption, such as the evaluation performance on how well medical BERT encodes CT radiology reports.
The performance evaluation of the whole pipeline model cannot provide direct evidence for the capability of the language model in the pipeline, because the evaluation is not thorough enough to identify potential biases and distribution shifts, and to capture the fine-grained capability differences of the language model on the target task for different conditioned pathologies. Also, the evaluation performance can be explained away by confounding factors in the pipeline.

It is good practice that the authors reported other important design assumptions in the method section that assume the inclusion of clinical prior knowledge of radiology report and anatomical structures in the modeling can guide image generation, and disclosed other detailed modeling assumptions in the limitation part, including modeling limitations on disease type and scale, and the accuracy limitation of anatomical segmentations.

\textbf{\ref{item:eval} Evaluation} 

\textbf{Computational evaluation.} 
Regarding the dataset, the authors used an internal dataset from a local hospital, but did not report whether research ethics board (REB) review and approval were needed to use the dataset. This is to ensure that the data usage process are ethical and ethics requirements, such as consent process, privacy protection, data usage, access, storage, retention and destruction, are met.

Regarding the realism evaluation, to calculate the perceptual metrics of FID and MMD on the distance between real and synthetic images, the authors utilized a 3D ResNet model pre-trained on eight 3D medical image segmentation datasets for feature extraction. 
As we analyzed in \cref{list:fm_assessor_validity} in \cref{llm}, this evaluation method, which employs a pre-trained model for automated realism assessment, lacks construct validity as it fails to ground the measurement of realism in the target medical knowledge and datasets. The intrinsic limitation of this evaluation method that uses a pre-trained model as the realism assessor should be acknowledged in the limitation section of the paper. The pre-trained ResNet model did not undergo evaluation on how well it learned the target features, therefore it is questionable whether the pre-trained ResNet model can be a valid representation for the target medical knowledge and datasets. The authors complemented the realism evaluation with radiologists' assessment, therefore avoiding using the pre-trained model as the sole assessor for realism evaluation. 

\textbf{User study evaluation.}
It is good practice that the authors conduced a user study with radiologists to assess MedSyn. For studies involving human subjects, it is standard research practice to obtain REB approval. However, the authors did not report whether the study underwent REB review and whether it was approved. Moreover, the participants' informed consent process was also not reported.

The authors failed to report important details in the study design, raising reasonable suspicion on the ethical conduct and scientific validity of the study. First, the recruitment details (location, duration, method, sampling process, etc.), and the inclusion and exclusion criteria were missing from the report. Since this user study is an online survey, it is unknown whether the participants were truly ``board-certified radiologists'' or not, raising doubt on the validity of results.
Second, deception was used in the study design: ``we intentionally did not mention to the radiologists that some of the CT scans they were about to interpret were generated by AI to avoid potentially biased results due to potentially negative perceptions towards AI.'' This may be unethical as it could pose more than minimal risks to participants (for example, inducing participants' false impression on the manifestation of lung CT), and should be determined by REB. The authors also did not mitigate the harms of deception, such as revealing to participants which images were synthetic (not real) at the end of the study.
Third, it is unclear how many image samples in total were included in the study and how they were sampled, 
which determines whether the samples were representative and the size provided enough statistical power.

\textbf{\ref{item:limit} Limitation analysis}

Similar to the analysis in Case 1, although the authors declared technique-specific limitations, they failed to declare the intrinsic limitations of any MISyn techniques, and did not conduct limitation assessment to identify the scope, weaknesses, and failure modes of MedSyn, such as the investigation on potential biases and distribution shifts introduced in the synthetic images. 

\section{Template for declaration of the intrinsic limits of an medical image synthesis technique} \label{app:statement}

Because the intrinsic limits of MISyn are shared by all MISyn techniques, 
to facilitate the declaration of intrinsic limits of MISyn in \ref{item:limit}, we provide a template that lists the intrinsic limits mentioned in this paper and their brief rationale. This template can be a starting point for authors to include in the MISyn papers and can be modified or expanded as necessary, e.g. if more intrinsic limits of MISyn are identified in other or future works.

This template is only for the declaration of the intrinsic limits. For other declarations such as the technical-specific limitations and design assumptions, we could not provide a template because these declarations are dependent on the specific MISyn technique.

\noindent \textbf{Template}

A medical image synthesis (MISyn) technique and the synthetic images it generates have intrinsic limits that cannot be overcame by more technical advancements. Because of the intrinsic limits, synthetic images can never fully replace the role of medical images. Therefore, the MISyn technique and the synthetic images it generates may not  be appropriate for use  in certain circumstances. Their appropriateness should be critically assessed, in conjunction with the exploration of alternative, non-synthetic image-based problem-solving approaches.

The intrinsic limits of the MISyn and its generated synthetic images include:

\begin{enumerate}[leftmargin=*]
\item Unlike medical images that are from patients to reflect the real-world medical phenomena, synthetic images do not automatically establish grounding in the real-world medical phenomena.

\item When training an MISyn-based model, synthetic images cannot fully resemble medical images, because the medical images and knowledge that  MISyn aims to model are complex phenomena and irreducible.

\item Because medical knowledge and datasets are irreducible to an MISyn model, compared to its training data, the MISyn modeling process inevitably introduces new distribution shifts and biases to the synthetic data. The distribution shifts and biases cannot be completely removed.

\end{enumerate}

In addition to the intrinsic limits of the MISyn model and its synthetic images, the evaluation method of MISyn includes the following intrinsic limits:

\begin{enumerate}[leftmargin=*]
\item The evaluation method is designed in a confined, experimental setting, which cannot fully assess the performance of the model in the open setting of the real world.

\item {[Optional, if a pre-trained model is used as the realism assessor for realism evaluation (assuming that other realism evaluation methods are also used that are grounded in the target medical knowledge and datasets)]} The evaluation method of using pre-trained model as the realism assessor has an intrinsic limit: the pre-trained model as realism assessor can only ground the realism evaluation in the pre-trained model space as an approximation, but cannot fully represent the target medical knowledge and datasets that the realism evaluation aims to assess.
\end{enumerate}

\small
\bibliographystyle{model2-names.bst}

\bibliography{ref}

\end{document}